# Cosmology in Flat Space-Time

*(Revised October 16, 2006)*

## ABSTRACT

A Lorentz-invariant cosmology based on E. A. Milne's Kinematic Relativity is shown to be capable of describing and accounting for all relativistic features of a world model without space-time curvature. It further implies the non-existence of black holes and the cosmological constant. The controversy over the value of the Hubble constant is resolved as is the recent conclusion that the universe's expansion is accelerating. "Dark matter" and "dark energy" are possibly identified and accounted for as well.

### Cosmology In Flat Space – Time

# Cosmology in Flat Space-Time

## 1. Absolute Space and Time

**C**osmology is the study of the entire universe as a physical system. Therefore in principle every physical application is a cosmological application. In a narrower usage, however, most of current cosmology concerns itself with world-wide or "global" problems such as model universes In such universes, one may frame specific special problems, such as the conditions or processes in the "early universe". In any case, these problems are subject to the constraints characteristic of the model universe chosen; the universes of general relativity and of the late steady state theory or of kinematic relativity are salient examples.

Let us illustrate some of the problems and methods characteristic of cosmology, at the same time making use of results already derived. To keep the discussion within tractable bounds, let us confine our attention to a universe which in its essential properties is as simple as is consistent with both interest and correspondence to the real world of physical observation. Exotic models will be left to treatises specifically devoted to more general cosmologies.

We begin, therefore, by appealing to Helmholtz' Theorem that **free mobility of rigid bodies of non-infinitesimal extension requires that space be Riemannian and of constant curvature.** The manifest existence and mobility of rigid bodies of more than infinitesimal extent makes this a highly desirable, if not necessary prerequisite; only if the more general consequences of such an assumption make it appear untenable would one discard it. Similarly, by Schur's Theorem, **a space is isotropic only if the curvature invariant R is constant.** Since the large-scale universe appears highly isotropic as observed in the primeval fireball radiation, this is a second cogent reason to seek a model universe in a Riemannian space of constant curvature. Inasmuch as the number of model universes of constant curvature is currently none, there is therefore also an element of novelty in exploring this possibility.

A model universe is, simply, a map in space-time. It specifies a density distribution defined through all space and for an infinite range of time. To possess credibility, it must also incorporate a law of gravitation compatible with its continued existence, present state and/or presumed history. It must further satisfy some form of relativity, so that it is not modified in its essential characteristics by an exchange of observers (spatial locations) or epochs (times).

To construct a map in space-time, it is first necessary to have a clock. Therefore the logical first step in recovering absolute time is to define a clock in some acceptable operational manner; traditional discussions simply assume "identical clocks" and "standard meter sticks". E. A. Milne was the first to define a *generalized clock* as a *monotonic single-valued parametrization of the sequence of events at a single observer*. Thus any observer O sees a succession of events **at himself** and monotonically assigns real numbers to those events according to the earlier-later ordering relation. These numbers are the "times" or epochs of the respective events at himself. (The observer does not at this stage presume to designate the epochs of events elsewhere.) This generalized clock may be given a conventional mechanical form with face and hands, which provide as events to parametrize the coincidences of the hands with a scale of numbers on the face.

Another observer $\bar{O}$ may do the same. The results will in general be entirely different, for the parametrizations are arbitrary. By what means can two arbitrary clocks be tested to determine whether or not they qualify as "identical (congruent) clocks" of traditional special relativity theory, and if they are found not to be identical, how may one or both be altered – **regraduated** – so that they will so qualify?

Milne used the reflexive property of clock equivalence to prescribe an operational test. Suppose $O$ sends a light signal to $\bar{O}$. The signal is sent forth when $O$'s clock reads $t_{(1)}$. When the signal arrives at $\bar{O}$, $\bar{O}$'s clock reads $\bar{t}_{(2)}$. By sending a modulated signal during some non-zero interval of time, observers $O$ and $\bar{O}$ can establish that

$$\bar{t}_{(2)} = \psi(t_{(1)}) \geq t_{(1)}.$$

If $\bar{O}$ returns the signal to $O$ by reflection, and if $O$ receives it back at time $t_{(3)}$ according to his own clock, they can further establish that

$$t_{(3)} = \bar{\psi}(\bar{t}_{(2)}) \geq \bar{t}_{(2)}.$$

If the clocks used by $O$ and $\bar{O}$ are "identical", it is by definition both necessary and sufficient that the functions $\psi$ and $\bar{\psi}$ be identical. Milne and Whitrow have shown (Zeitschrift für Astrophysik, v. 15, p. 270 (1938)) that it is always possible to find a transformation of one or the other of the clocks so that the clocks of $O$ and $\bar{O}$ may be made congruent in an operationally precise sense.

With light signals and a clock, each observer can assign space-time coordinates to any observable event $E$. Such coordinates are conventional to the extent that their definition is at the discretion of the observer at $O$. The convention which offers the appeal of greatest simplicity and the closest correspondence to everyday experience is the radar-ranging convention (adopted by Milne before the advent of radar) that a distant event $E$ occurs at a time

$$t'_{(2)} = \frac{1}{2}(t_{(3)} + t_{(1)})$$

and at a distance

$$r'_{(2)} = \frac{1}{2}c(t_{(3)} - t_{(1)}),$$

where $t_{(1)}$ and $t_{(3)}$ are, respectively, the times of departure and return of a light signal from $O$ reflected at the distant event. Note that in general $t'_{(2)} \neq \bar{t}_{(2)}$; the prime distinguishes the time ascribed by $O$ to the arrival of the signal at $E$ from the time $\bar{t}_{(2)}$ kept by the clock at $\bar{O}$. The difference is the aforementioned "relativity of time". The remaining two coordinates are position angles, measurable at each observer in the standard way. The constant $c$ is the velocity of light.

Exchange of observers is achieved in a mathematical (though of course, not physical) sense by a transformation of coordinates. We therefore seek a coordinate transformation in a space of constant curvature. The simplest such space is one of zero curvature, a "flat" 4-dimensional space-time. The Lorentz transformation between observers moving radially with respect to one another at a constant velocity $V$ represents a transformation which preserves the constancy of the velocity of light, a *physical* (i.e., *experimental*) requirement which the mathematics must reproduce.

A very important feature of the Lorentz transformation is that it is not, as heretofore, a transformation of the coordinates of an event $E$ from one system of coordinates at a given point $O$ in space-time to a different set of coordinates *at the same point* $O$. It is rather a transformation from the coordinates of $E$ at $O$ to the coordinates of the same event $E$ at some other point $\overline{O}$ where $\overline{x}^i = \overline{x}^i(x^j)$, ( $i, j = 1, 2, 3, 4$). Recall that observer $\overline{O}$ coincided with observer $O$ at time $t = 0 = \overline{t}$ and has since been moving along the common $x$ – (or $\overline{x}$ – ) axis at constant speed $V$.

We must therefore here introduce a practical matter which has not arisen in our discussions prior to this point: How are the coordinates $x^\mu$ and $\overline{x}^\mu$ of an event $E$ determined? Since the Lorentz transformation was derived to ensure that the velocity of light would be the same to observers $O$ and $\overline{O}$ in uniform relative (radial!) motion, the velocity of light provides a standard by which the ratio of spatial and temporal lengths may be compared in the experience and observations of $O$ and $\overline{O}$; it is the common denominator between observers who may otherwise be inaccessible, as would be the case with an observer $O$ on the earth and an observer $\overline{O}$ in the Andromeda galaxy.

If $c$, the velocity of light, can be relied upon to be a constant of both observers' experience, then a radar procedure would surely be as simple and as direct a means of measuring distances as could be devised. Thus, for $O$ to determine $\overline{O}$'s distance, he could send a modulated signal to $\overline{O}$, arranging that it be returned by reflection. If it were sent at a time $t_{(1)}$ and received back at $t_{(3)}$, and if the signal traveled at the same speed both outward and back, it must be true that

$$t'_{(2)} - t_{(1)} = t_{(3)} - t'_{(2)},$$

where $t'_{(2)}$ is the time which $O$ ascribes to its arrival at $\overline{O}$ . Note that $t'_{(2)}$ is an *inferred* time (hence the distinguishing prime), since no meaning could be attached to an assertion that the signal arrived at $\overline{O}$ at a time $t'_{(2)}$ as indicated on $O$'s clock. By re-arranging the preceding equation, $O$ ascribes the signal's arrival at $\overline{O}$ (or any point not coincident with $O$, for that matter) as at time

(1.1)                     $$t'_{(2)} = \frac{1}{2}\Big(t_{(3)} + t_{(1)}\Big).$$

Observer $O$ will, by the radar technique, also ascribe to $\bar{O}$ a distance equal to

$$r'_{(2)} = c\,(\,t'_{(2)} - t_{(1)}\,) = c\,(\,t_{(3)} - t'_{(2)}\,),$$

the first being the distance the beam of light travels on its outward journey and the second the distance it travels on its return. Substituting the expression for $t'_2$ into either one, we get

(1.2) $$r'_{(2)} = \frac{1}{2}\,c\,(\,t_{(3)} - t_{(1)}\,).$$

Since the symmetry between $O$ and $\bar{O}$ demands that they may be interchanged without effect, it must also be true that

$$\bar{t}'_2 = \frac{1}{2}\,(\,\bar{t}_3 + \bar{t}_1\,) \;\; \text{and} \;\; \bar{r}'_2 = \frac{1}{2}\,c\,(\,\bar{t}_3 - \bar{t}_1\,),$$

where the barred variables have meanings for $\bar{O}$ corresponding to the same unbarred variables' meanings to $O$. Both observers, therefore, can make observations of the other to determine the other's location at a particular time, hence the other's relative motion. It is not a foregone conclusion, however, that at the instant $O$'s light signal is reflected off

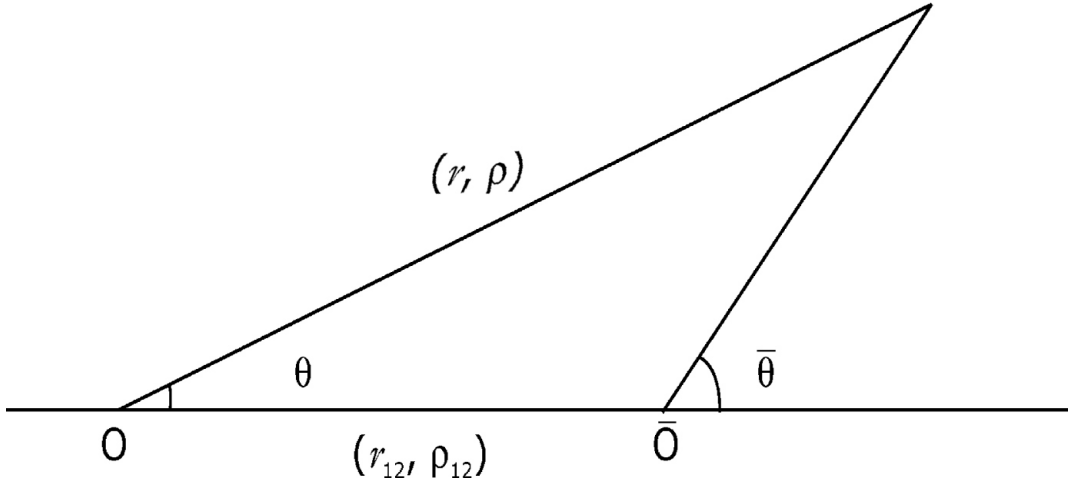

$$E\,(\,r, t:\; \bar{r},\; \bar{t}\,)$$

$$(\,r,\; \rho\,)$$

$$\theta \qquad \bar{\theta}$$

$$O \qquad (\,r_{12},\; \rho_{12}\,) \qquad \bar{O}$$

Figure 94

$\bar{O}$'s clock, the clock at $\bar{O}$ will read $t'_{(2)}$. In fact, if $t'_{(2)}$ and $r'_{(2)}$ are substituted into the Lorentz transformation, we realize that they will *not* be the same. By having used a modulated signal, however, we can determine that $\bar{t}_{(2)}$ is a function $\bar{t}_{(2)} = f(\,t'_{(2)}\,)$ of the ascribed time $t'_{(2)}$. Reciprocity then demands that $t_{(2)} = f(\,\bar{t}'_{(2)}\,)$, where the function $f$ is identically the same *function* in both cases. Under the Lorentz transformation, that function is

$$f(x) = \left(\sqrt{1 - V^2/c^2}\right)x \;\; \text{or} \;\; \bar{t} = \sqrt{1 - V^2/c^2}\,t'.$$

The clocks of $O$ and $\bar{O}$ are then said to be **congruent.** If they were not so originally, then one or the other of $O$ or $\bar{O}$ may adopt a clock whose readings do satisfy the required relation. This is a process called **regraduation.**

**Let us generalize the problem one step further. Suppose that another** observer $O^*$ is also present. If $\bar{O}$ and $O^*$ regradute their clocks so that their respective clocks are congruent with $O$'s, will they then also be congruent with each other? E. A. Milne showed that they will not be mutually congruent unless (a) all three observers coincided at time $t = \bar{t} = t^* = 0$ and (b) they are moving radially away from $O$ (and hence from each other) at constant velocity. Such a set of observers (even infinitely many) constitute what Milne defined as a **kinematic equivalence.**[*]

Though we have referred to "observers" at $O$, $\bar{O}$, $O^*$, ..., etc., we may just as well substitute disembodied frames of reference. An equivalence thus provides at every point of space-time an inertial frame of reference and a clock, which in combination may be used to locate events elsewhere in a way which can be translated straightforwardly into the location of the same event from all other frames of reference of the equivalence. Clearly, this is essential to any model universe or **world model.**

The equivalence in which various observers' clocks are related by the Lorentz transformation is one in which those observers are separating radially from one another at constant speed $V = r/t$. It therefore represents an "expanding universe" in exact correspondence with the observed universe which obeys a law of redshifts that implies uniform expansion at velocity $V = r/t$. It is gratifying that such a model expanding universe mirrors one of the most distinctive features of the observed universe, a feature wholly unanticipated prior to its discovery by Edwin Hubble in the 1920s.

On the other hand, there appears to be no way to graft onto the uniformly expanding universe universal laws of motion and of gravitation. This untoward circumstance comes about because neither $O$'s times $t$ nor distances $r$ are Lorentz invariants. This implies that observations made by any other observer $\bar{O}$ of the equivalence and expressed in his own coordinates $\bar{t}$ and $\bar{r}$ will describe events in a way which depends, through the Lorentz transformation, upon the specific identities of $O$ and $\bar{O}$; that is to say, laws of motion and gravitation derived inductively by $O$ and $\bar{O}$ must include, at least implicitly, the identity of the observer and will therefore not be universal. On its face, this objection appears to be a fatal one to the expanding equivalence model of the universe.

At the same time, knowing that the root of the difficulty lies in the fact that neither $t$ nor $r$ is a Lorentz invariant suggests where to look for a solution. We have already seen that $\tau = \sqrt{t^2 - r^2/c^2} = t\sqrt{1 - r^2/(ct)^2}$ is a Lorentz invariant of the dimensions of time. Moreover, when $r \ll ct$, the *value* of $\tau$ differs from the *value* of $t$ by only a very small amount which is usually far below the possibilities of observation.

We are therefore prompted to look for a transformation to coordinates in which $\tau$ will replace $t$. At the same time we must replace $r$ with some $\rho$ which is also Lorentz invariant. This will make it possible for any observer $O$ to translate his own observations

---

[*]E. A. Milne, "*Kinematic Relativity*", (Oxford (1948)), Ch. II, §§14, 15, pp. 22-24

of an event $E$ (or continuum of events, a **history**) in such a way that it agrees with the same event or history by any other observer $\overline{O}$ of the congruence. The requirement of universality will then have been met.

Let us seek to give it a simple common-sense derivation. In reviewing our method of determining how to establish congruent clocks for observers $O$ and $\overline{O}$, let us recall that $O$ assigned to an event $E$ a time $t'_{(2)}$ which was the arithmetic mean of sending time $t_{(1)}$ and return time $t_{(3)}$ of a light signal; this was suggested by the requirement that the velocity of light be constant in empty space. Thus, $t'_{(2)} = \frac{1}{2}(t_{(3)} + t_{(1)})$. At the same time, the quantity

$$\tau'_{(2)} = \left[\left(t'_{(2)}\right)^2 - \left(\frac{r'_{(2)}}{c}\right)^2\right]^{1/2} = \left[\left(t'_{(2)} + \frac{r'_{(2)}}{c}\right)\left(t'_{(2)} - \frac{r'_{(2)}}{c}\right)\right]^{1/2} = \sqrt{t_{(3)}\, t_{(1)}}$$

is an invariant time-like quantity which, however, does not have the form of the arithmetic mean of $t_{(1)}$ and $t_{(3)}$. Instead, it is the geometric mean of $t_{(1)}$ and $t_{(3)}$. Now it is well known that the logarithm of the geometric mean of $A$ and $B$ is the arithmetic mean of the logarithms of $A$ and $B$. Therefore let us write

$$\frac{\tau'_{(2)}}{t_{(0)}} = \left[\left(\frac{t_{(3)}}{t_{(0)}}\right)\left(\frac{t_{(1)}}{t_{(0)}}\right)\right]^{1/2},$$

where $t_{(0)}$ is some convenient constant of the dimensions of time, introduced to render the three ratios dimensionless numbers. Then

$$\ln\left(\frac{\tau'_{(2)}}{t_{(0)}}\right) = \ln\left[\left(\frac{t_{(3)}}{t_{(0)}}\right)\left(\frac{t_{(1)}}{t_{(0)}}\right)\right]^{1/2} = \frac{1}{2}\left[\ln\left(\frac{t_{(3)}}{t_{(0)}}\right) + \ln\left(\frac{t_{(1)}}{t_{(0)}}\right)\right].$$

To make each of these terms time-like, let us multiply throughout by a second dimensional constant $t^*$. Then

$$t^* \ln\left(\frac{\tau'_{(2)}}{t_0}\right) = t^*\left\{\frac{1}{2}\left[\ln\left(\frac{t_{(3)}}{t_0}\right) + \ln\left(\frac{t_{(1)}}{t_0}\right)\right]\right\}.$$

If this equation is to parallel the equation $t'_{(2)} = \frac{1}{2}\left(t_{(3)} + t_{(1)}\right)$, it is clear that we should choose a new time variable

$$T = t^* \ln\frac{\tau}{t_{(0)}},$$

for then the previous equation becomes

$$T'_{(2)} = \frac{1}{2}\left(T_{(3)} + T_{(1)}\right),$$

wherein the ascribed time is a Lorentz invariant; all observers of the equivalence will agree upon the reflection time $T'_{(2)}$.

Two final details may be added, purely for reasons of esthetics and economy. First, let us arrange that $T = t_{(0)}$ when $\tau = t_{(0)}$. This requires that we add $t_{(0)}$ to the right hand side:

$$T = \left( t^* \ln \frac{\tau}{t_{(0)}} \right) + t_{(0)}.$$

Secondly, let us arrange also that $d T = d\tau$ at $T = \tau = t_{(0)}$. This requires that

$$d T = t^* \frac{d\tau}{\tau} = t_{(0)} \frac{d\tau}{\tau} \text{ at } T = \tau = t_{(0)},$$

or $t^* = t_{(0)}$. With these conventions, the final regraduation of the time variable becomes

$$(1.3) \qquad\qquad T = t_{(0)} \left( 1 + \ln \frac{\tau}{t_{(0)}} \right),$$

with inverse

$$(1.4) \qquad\qquad \tau = t_{(0)} e^{(T - t_{(0)})/t_{(0)}}.$$

Note that this regraduation is a *tensor transformation*, one which is made by each observer to change to a new time coordinate at himself, irrespective of what other observers may do.

This suggests that $O$, $\overline{O}$, $O^*$, $\ldots$, adopt new clocks whose times $T$ are related to the former times $\tau$ by Eq. (1.4) or its inverse Eq. (1.3). If *all* observers of the equivalence make such a regraduation, their clocks will still be congruent because their original clocks were. The kinematic equivalence now becomes a **dynamical equivalence**.

The kinematic time $t$ is measured from the moment when all observers $O$, $\overline{O}$, $\ldots$, were coincident, and therefore ranges from $t = \overline{t} = \ldots = 0$ to infinity. Since $\tau = \sqrt{t^2 - r^2/c^2} \leq t$, the range of $\tau$ is also $(0, +\infty)$. By Eq. (1.3), therefore, the range of the time $T$ is $(-\infty, +\infty)$.

Let us now calculate the coordinates of a distant event $E$ (see Fig. 94). If the light signal sent out at time $T_{(1)}$ returns at time $T_{(3)}$, observer $O$ imputes to the event $E$ a distance

$$r'_{(2)} = \frac{1}{2} c \left( t_{(3)} - t_{(1)} \right) = \frac{1}{2} c \, t_{(0)} \left[ e^{(T_{(3)} - t_{(0)})/t_{(0)}} - e^{(T_{(1)} - t_{(0)})/t_{(0)}} \right].$$

However, it is also true that

$$T_{(3)} = T_{(2)} + \frac{\rho_{(2)}}{c} \text{ and } T_{(1)} = T_{(2)} - \frac{\rho_{(2)}}{c},$$

where $\rho_{(2)}$ is the new radial coordinate of the event $E$ observed at time $T_{(2)}$. We see then that

$$r'_{(2)} = c\, t_{(0)}\, e^{\left(\frac{T_{(2)} - t_{(0)}}{t_{(0)}}\right)} \left[\frac{e^{\left(\rho_{(2)}/ct_{(0)}\right)} - e^{-\left(\rho_{(2)}/ct_{(0)}\right)}}{2}\right] = c\, \tau_{(2)} \sinh \frac{\rho_{(2)}}{ct_{(0)}} \; .$$

Dropping the common subscript, we have

$$(1.5) \qquad\qquad\qquad r = c\, \tau \sinh \frac{\rho}{c\, t_{(0)}} \; .$$

Similarly, the time ascribed to the event $E$ is

$$t'_{(2)} = \frac{1}{2}\left(t_{(3)} + t_{(1)}\right) = t_{(0)}\, e^{(T_{(2)} - t_{(0)})/t_{(0)}} \left[\frac{e^{\left(\rho_{(2)}/ct_{(0)}\right)} + e^{-\left(\rho_{(2)}/ct_{(0)}\right)}}{2}\right] = \tau_{(2)} \cosh \frac{\rho_{(2)}}{ct_{(0)}} \; .$$

Again, by dropping needless subscripts and primes,

$$(1.6) \qquad\qquad\qquad t = \tau \cosh \frac{\rho}{c\, t_{(0)}} \; .$$

The final forms of the desired transformation are therefore

$$t = \tau \cosh \frac{\rho}{ct_{(0)}}, \quad T = t_{(0)}\left[1 + \frac{1}{2}\ln \frac{t^2 - r^2/c^2}{t_{(0)}^2}\right],$$

$$(1.7)$$

$$r = \tau \sinh \frac{\rho}{ct_{(0)}}, \quad \rho = \frac{1}{2}\, c\, t_{(0)} \ln \frac{t + \dfrac{r}{c}}{t - \dfrac{r}{c}} \; .$$

The associated world-line elements are

$$(1.8)$$

$$(ds)^2 = (d\tau)^2 - \left(\frac{\tau}{ct_{(0)}}\right)^2 \left\{(d\rho)^2 + \left(ct_{(0)}\right)^2 \sinh^2 \frac{\rho}{ct_{(0)}}\left[\cos^2\varphi\,(d\theta)^2 + (d\varphi)^2\right]\right\}$$

or, in the $(T, \rho)$ - coordinates,

$$(1.9)$$

$$\tau^{-2}(ds)^2 = (dT)^2 - \frac{1}{c^2}\left\{(d\rho)^2 + (ct_{(0)} \sinh \frac{\rho}{ct_{(0)}})^2 \left[\cos^2\varphi\,(d\theta)^2 + (d\varphi)^2\right]\right\} \; .$$

From the expressions for the square of the line element in $(r, t)$-**coordinates** and in $(\rho, T)$-**coordinates** it is clear that the transformation between them constitutes a conformal mapping. The overall curvature of the 4-space with this metric is zero, for we have transformed from Euclidean coordinates $(r, t)$ to dynamical variables $(\rho, T)$ and a mere transformation of coordinates cannot change the curvature of the space. However, the spatial cross-section $T = t_0$ has a constant negative curvature $K = -(ct_0)^{-2}$. (This is analogous to the fact that in ordinary spherical coordinates the sub-space defined by $r = r_0$ has a positive curvature $K = r_0^{-2}$). The "radius of curvature" of the 3-dimensional sub-space $T = t_0$ is therefore $R = ct_0$.

**The constant $t_{(0)}$ is thus far merely an arbitrary constant. How ought it be** chosen? Since the entire purpose of this development is to find a flat-space representation of the universe, perhaps a model universe can suggest a value for $t_{(0)}$. Though there is no logical compulsion for one value rather than another, there may be a pragmatic advantage in some particular choice.

Such is the case. Recall that in the expanding model of the universe we have $V = r/t$ where $V$ is the (constant) recessional velocity of a galaxy at observed distance $r$. From this we find a time $t = r/V$ which has nearly the same value *for all the nearer galaxies* — i.e., a "universal" constant. Let us identify this value with $t_{(0)}$. Observations then presently give for $t_{(0)}$ the value $t_{(0)} \approx 13 \times 10^9$ years. If we adopt this value, we see that $K = -1/(ct_{(0)})^2$ is approximately $6 \times 10^{-21} (\text{light years})^{-2}$, an extremely small quantity on all but a cosmological scale. We see also that at the present time, times and distances are $t \approx T$, $\rho \approx r$ even on a scale up to many (millions!) light years. At the same time, the world-line element becomes very nearly

$$(ds)^2 = (dt)^2 - \frac{1}{c^2}\left\{(dr)^2 + r^2\left[\cos^2\varphi\,(d\theta)^2 + (d\varphi)^2\right]\right\},$$

the form it would have had in Euclidean space.

We have in this way found time and distance coordinates such that all observers in the universe who are related by the same Lorentz transformation (aside from the value of $V$) will agree upon the times and locations of events anywhere in the universe. In this sense, we have constructed "absolute space-time".

It is of interest to see what the Lorentz transformation becomes in this absolute space-time. In Cartesian coordinates $x$, $y$, $z$, and $t$ the transformation of $t$ to $\bar{t}$ is

$$\bar{t} = \frac{t - \dfrac{Vx}{c^2}}{\left[1 - \dfrac{V^2}{c^2}\right]^{1/2}}.$$

Making the substitutions for old and new coordinates, and using the relation that

$$(1.10) \qquad V = \frac{r}{t} = \frac{c\tau \sinh\dfrac{\rho}{ct_{(0)}}}{\tau \cosh\dfrac{\rho}{ct_{(0)}}} = c\tanh\frac{\rho}{ct_{(0)}},$$

the Lorentz transformation for $\bar{t}$ is

$$\bar{t} = \bar{\tau} \cosh \frac{\bar{\rho}}{c\, t_{(0)}} = \frac{\tau \cosh \dfrac{\rho}{c\, t_{(0)}} - \tau \tanh \dfrac{\rho_{(12)}}{c\, t_{(0)}} \sinh \dfrac{\rho}{c\, t_{(0)}} \cos \theta}{\left[ 1 - \tanh^2 \dfrac{\rho_{(12)}}{c\, t_{(0)}} \right]^{1/2}}$$

$$= \tau \left[ \cosh \frac{\rho}{c\, t_{(0)}} \cosh \frac{\rho_{(12)}}{c\, t_{(0)}} - \sinh \frac{\rho}{c\, t_{(0)}} \sinh \frac{\rho_{(12)}}{c\, t_{(0)}} \cos \theta \right].$$

But the triangle $\mathrm{O\bar{O}E}$ (Fig. 94) is a geodesic triangle. Therefore, since by the cosine law of hyperbolic geometry,

$$\cosh \frac{\bar{\rho}}{c\, t_{(0)}} = \cosh \frac{\rho}{c\, t_{(0)}} \cosh \frac{\rho_{(12)}}{c\, t_{(0)}} - \sinh \frac{\rho}{c\, t_{(0)}} \sinh \frac{\rho_{(12)}}{c\, t_{(0)}} \cos \theta \,,$$

the first of the Lorentz transformation equations reduces to $\tau = \bar{\tau}$, the contemporaneity condition, stating merely that $\mathrm{O}$ and $\mathrm{\bar{O}}$ observe the same event $\mathrm{E}$.

The second equation of the Lorentz transformations is

$$\bar{x} = \left( x - \mathrm{V} t \right) \left[ 1 - \frac{\mathrm{V}^2}{c^2} \right]^{-1/2},$$

which becomes

$$\bar{r} \cos \bar{\theta} = \frac{r \cos \theta - \mathrm{V} t}{\left[ 1 - \dfrac{\mathrm{V}^2}{c^2} \right]^{1/2}}.$$

When substitution is made for $\bar{r}$, $r$, $\mathrm{V}$ and $t$, and when $\cos \bar{\theta}$ and $\cos \theta$ are eliminated by means of the same cosine law of hyperbolic trigonometry, this second of the Lorentz transformation equations again reduces to the simultaneity condition $\tau = \bar{\tau}$. The third and fourth Lorentz transformation equations, using the sine law of hyperbolic trigonometry again leads to the condition $\tau = \bar{\tau}$. We may therefore say that **the Lorentz transformation is a transformation of the coordinates of an event $\mathrm{E}$ from the frame of an observer $\mathrm{O}$ to the frame of an observer $\mathrm{\bar{O}}$ in absolute space at the same absolute time.**

It is interesting that in the "expanding" universe there exists a velocity-distance relation which says that $r / t = \mathrm{V}$. Thus more distant frames of reference (greater $r$) have proportionately greater recessional velocities $\mathrm{V}$. In absolute space-time, however, the velocity-distance relation becomes

(1.11) $$\mathrm{V} = c \tanh \frac{\rho}{c\, t_{(0)}}, \quad \rho = c\, t_{(0)} \ln \left[ \frac{1 + \mathrm{V}/c}{1 - \mathrm{V}/c} \right]^{1/2}.$$

For fixed $\mathrm{V}$, $\rho$ has a value dependent only upon the age of the universe, $t_0$.

## 2. Gravitation in Flat Space-Time

**A** model universe will be dismissed unceremoniously if it does not offer some theory of gravitation. We may therefore ask: Can a theory of gravitation be advanced by stating equations in absolute space and time which will pass the crucial tests of (1) the advance of Mercury's perihelion, (2) gravitational refraction at the limb of the sun, (3) the gravitational red shift, (4) the echo delay of sun-grazing light beams, (5) the radiation of gravitational energy, (6) objects of very large mass at the centers of all galaxies and large globulars, and (7) give prospect of accounting for "dark energy" and "dark matter".

**(1) Consider a mass M, such as the sun, in an otherwise empty universe. Now let** a mass $m$ (such as a planet) be brought from infinity along any path whatsoever in a quasistatic manner (just as infinitely slow changes are effected in quasistatic thermodynamic processes). According to Newton's classical law of gravitation, an amount of work

$$d\mathrm{W} = \mathrm{G}\frac{\mathrm{M}m}{r^2}(-dr)$$

is done on the mass $m$ during the displacement $-dr$. (In a concession to time-honored notation, we have reverted to the use of the radius vector $r$ rather than the absolute distance $\rho$). According to special relativity, mass and energy are equivalent and inter-convertible. If we therefore assume that the work done by gravitation appears as an increase of mass, we have

$$d\mathrm{W} = c^2\,dm = -\frac{\mathrm{GM}}{r^2}m\,dr, \quad \frac{dm}{m} = -\mathrm{G}\frac{\mathrm{M}}{c^2}\frac{dr}{r^2},$$

(2.1) $$\ln\frac{m}{m_0} = \frac{\mathrm{GM}}{c^2 r}, \quad m(r) = m_0\,e^{\mathrm{GM}/c^2 r},$$

where $m_0$ is the value of $m(r)$ at $r = \infty$. (What is true for mass $m(r)$ is also true for mass $\mathrm{M}(r)$ because the relation of $m$ and $\mathrm{M}$ is entirely symmetric. Therefore we should strictly write

$$\mathrm{M}(r) = \mathrm{M}_0\,e^{\mathrm{G}(m_0+\mathrm{M}_0)/c^2 r} \text{ and } m(r) = m_0\,e^{\mathrm{G}(\mathrm{M}_0+m_0)/c^2 r}$$

but since for the earth $m_0 = .00003\,\mathrm{M}$, we omit the mass of the earth.)

By analogy with classical dynamics, let us now define the energy-momentum 4-vector as

$$p^\alpha = c\,m(r)\,\mathrm{v}^\alpha = c\,m_0\,\mathrm{v}^\alpha e^{\mu/c^2 r}, \quad \mu = \mathrm{GM}.$$

We then calculate $1/2\,m_0$ times the square of the invariant magnitude of the energy-momentum vector to be

$$\frac{1}{2\,m_0}\left|p^\gamma\right|^2 = \frac{g_{\alpha\beta}p^\alpha p^\beta}{2\,m_0} = \frac{m_0}{2}e^{2\mu/c^2 r}\{c^2 - \mathrm{v}^2\}e^{2(\mathrm{T}-t_0/t_0)}.$$

Consider the two terms separately. The second is the negative of the kinetic energy $K$ to the zeroth order in $c^2$. The first is

$$\frac{1}{2}\, m_0 \left[ c^2\, e^{\,2\,\mu/c^2 r} \right] e^{\,2\,(T - t_0)/t_0} = \left[ \frac{1}{2}\, m_0\, c^2 + \frac{m_0\, \mu}{r} + \dots \right] e^{\,2\,(T - t_0)/t_0}.$$

At the present time ($T = t_0$), the final term (outside the brackets) is 1. The quantity within the brackets is therefore clearly the negative of the potential energy $U$. Hence by choosing $\left| p^{\,\gamma} \right|^2 = \frac{1}{2}\, m_0\, c^2 - E$ we have the equation of the conservation of energy,

$$E = U + K.$$

Having identified the kinetic energy $K$ and potential energy $U$ and replacing $e^{\,2\,(T - t_0)/t_0}$ by 1, we can specify the Lagrangian[*] as

(2.2) $$L = \frac{1}{2}\, m_0 \left\{ \left( c^2 + v^2 \right) e^{\,2\,\mu/c^2 r} - c^2 \right\}.$$

With the above choice of expressions for the kinetic and potential energies, Lagrange's equations are then

(2.3) $$\ddot{r} - r\,\dot{\theta}^2 - \frac{\partial}{\partial r}\left( \frac{\mu}{r} \right) + \frac{1}{c^2}\left[ -v^2\, \frac{\partial}{\partial r}\left( \frac{\mu}{r} \right) + 2\,\dot{r}\, \frac{d}{dt}\left( \frac{\mu}{r} \right) \right] = 0,$$

and

(2.4) $$\frac{d}{dt}\left( r^2\, \dot{\theta}\, e^{\,2\,\mu/c^2 r} \right) = 0.$$

From this last equation we immediately obtain the integral

(2.5) $$r^2\, \dot{\theta}\, e^{\,2\,\mu/c^2 r} = h = \text{constant}.$$

---

[*] The theory of general relativity solves the two-body problem (one large mass $M$ and one small mass $m$) by finding the geodesics of the 4-metric

$$(ds)^2 = \left( 1 - \frac{2\,G M}{c^2 r} \right) (dt)^2$$
$$- \frac{1}{c^2}\left( \frac{(dr^2)}{1 - \dfrac{2\,G M}{c^2 r}} + r^2 \left[ \cos^2 \varphi\, (d\theta)^2 + (d\varphi)^2 \right] \right),$$

the celebrated Schwarzschild metric. Clearly, it has a singularity at $r = \dfrac{2\,G M}{c^2}$, the Schwarzschild radius, where the first term on the right hand side vanishes and the second term becomes infinite. This implies that the volume within the Schwarzschild radius is inaccessible from without along a geodesic. For a solar mass, the Schwarzschild radius is 1.80 miles or 2.95 kilometers.

Let equation (2.3) be re-cast into the form

$$(2.6) \qquad \ddot{r} \, - \, r \, \dot{\theta}^2 \, + \, \frac{\mu}{r^2} \left\{ 1 \, + \, \frac{r^2 \, \dot{\theta}^2 \, - \, \dot{r}^2}{c^2} \right\} \, = \, 0 \, .$$

The standard method of solution is to transform to a new dependent variable $u \, = \, 1 \, / \, r$ and use equation (2.5) to eliminate $dt$ in favor of $d\theta$. Then

$$(2.7) \qquad u'' \, + \, u \, + \, \frac{\mu}{h^2} \, + \, \frac{\mu}{c^2} \left\{ 4 \, u \left( u'' \, + \, u \right) \, + \, u^2 \, + \, u'^2 \right\} \, = \, 0 \, ,$$

subject to the initial conditions

$$(2.8) \qquad u(0) \, = \, \frac{\mu \, (1 \, + \, e)}{h^2} \, = \, \frac{1}{a \, (1 \, - \, e)} \, , \quad u'(0) \, = \, 0 \, .$$

(Primes indicate derivatives with respect to $\theta$.) The solution of equation (2.7), to the first order in $1/c^2$ is

$$(2.9) \qquad u(\theta) \, = \, \frac{\mu}{h^2} \left\{ 1 \, + \, e \cos \left[ (1 \, - \, \kappa) \, \theta \right] \, + \, \kappa (5 \, + \, e) \, (1 \, - \, \cos \theta) \right\} \, ,$$

where

$$(2.10) \qquad \kappa \, = \, 3 \left( \frac{\mu}{ch} \right)^2 \, .$$

Hence, when $\theta \, = \, 2 \, \pi$, $u(\theta) \, \neq \, u(0)$ because $(1 \, - \, \kappa) \, \theta$ is less than $2 \, \pi$ by an amount

$$(2.11) \qquad \Delta \, \theta \, = \, 2 \, \pi \, \kappa \, = \, 6 \, \pi \left( \frac{\mu}{ch} \right)^2 \, = \, \frac{24 \, \pi^3 \, a^2}{c^2 \left( 1 \, - \, e^2 \right) P^2} \, .$$

The perihelion of Mercury thus advances $\Delta \, \theta$ per revolution or 43" per century.

*Ex. (2.1) Geometrize the world-line element given in Eq. (1.9) and compare it to the Schwarzschild line element.*

*Ans. By Eq. (4.4.3),* $\overline{g}_{ij} \, = \, (E \, - \, V) \, g_{ij} \, = \, e^{2 \, \mu / c^2 \, r} \, g_{ij} \, .$

**(2) To determine the bending of light at the limb of the sun, multiply Eq. (2.3) by $2\dot{r}$ and Eq. (2.4) by $2 \, \dot{\theta}$ and add. Then**

$$(2.12) \qquad \frac{d \left( v^2 \right)}{dt} \, + \, \frac{2 \, \mu \, \dot{r}}{r^2} \left( 1 \, - \, \frac{v^2}{c^2} \right) \, = \, 0 \, .$$

Hence when $v^2 \, = \, c^2$, $v^2$ must be constant. To determine the trajectory of the photon, set $v \, = \, c$ in Eq. (2.3), whence

$$(2.13) \qquad \ddot{r} \, - \, r \, \dot{\theta}^2 \, + \, \frac{2 \, \mu}{r^2} \left( 1 \, - \, \frac{\dot{r}^2}{c^2} \right) \, = \, 0$$

or, since $c^2 = \dot{r}^2 + r^2 \dot{\theta}^2$

$$(2.14) \qquad \ddot{r} - \left( r - \frac{2\mu}{c^2} \right) \dot{\theta}^2 = 0 \, .$$

Eliminating $\dot{\theta}$ by means of Eq. (2.5), and expressing $\ddot{r}$ in terms of $u$ and its derivatives with respect to $\theta$, we get

$$(2.15) \qquad u'' + u = \frac{2\mu}{c^2} \left( u^2 + u'^2 \right) ,$$

subject to initial conditions $u(0) = 1/R$, $u'(0) = 0$, where $R$ is the photon's perihelion distance. The solution then is

$$(2.16) \qquad u(\theta) = \frac{2\mu}{c^2 R^2} + \left( 1 - \frac{2\mu}{c^2 R} \right) \frac{\cos\theta}{R} \, .$$

This is very nearly the equation of a hyperbola whose asymptotes are mutually inclined by the angle

$$2\varepsilon = 2\sin^{-1}\left[ \frac{2\mu}{c^2 R} \right] \approx \frac{4\mu}{c^2 R} \, ,$$

the required bending at the limb of the sun.

**(3) To obtain the gravitational red shift, we begin with equation (2.12), which** may be re-written in the form

$$\frac{1}{2} c^2 \frac{d\left( v^2/c^2 \right)}{1 - \left( v^2/c^2 \right)} = - \frac{\mu \, dr}{r^2} \, ,$$

$$(2.17) \qquad -\frac{1}{2}\ln\left( 1 - v^2/c^2 \right) = \frac{\mu}{c^2 r} + \frac{E}{m c^2} \, ,$$

$$\frac{m_0 c^2}{\left[ 1 - v^2/c^2 \right]^{1/2}} = m_0 c^2 \, e^{(E - U)/m_0 c^2} \, , \quad U = - \frac{\mu}{c^2 r}$$

where $E$, a constant of integration, is the total energy. In the classical limit as $c \to \infty$ equation (2.17) becomes

$$K = \frac{1}{2} m_0 v^2 = E + \frac{\mu m}{r} = E - U \, ,$$

the equation of conservation of energy (see Ex. (4.3.15)).

To apply Eq. (2.17) to a photon, consider the left and right hand sides separately. The left hand side, the kinetic energy of a particle, in the limit as $v \to c$ and $m_0 \to 0$, simultaneously, becomes $h\nu$, the kinetic energy of a photon. The right hand side may be given the form $h\,\nu_0\,e^{\mu/c^2 r}$ for an appropriately chosen $\nu_0$. On both sides of the equation, $h$ is here Planck's constant (not to be confused with the areal velocity of Eq. (2.5)). For a photon, therefore, Eq. (2.17) becomes

$$(2.18) \qquad\qquad \nu = \nu_0\,e^{\mu/c^2 r}.$$

The significance of $\nu_0$ is now clearly that it is the value of $\nu$ at $r = \infty$ or in the absence of gravitation ($\mu = 0$). The usual redshift equation follows by taking the logarithmic derivative of both sides, giving

$$(2.19) \qquad\qquad \frac{\Delta\nu}{\nu} = -\,\frac{\mu\,\Delta r}{c^2\,r^2}.$$

An interesting corollary of Eq. (2.18) is that for *any* positive $\nu$ and $r$, it follows that $\nu_0 > 0$. Therefore a photon would be able to escape even from a very compact finite mass, no matter how great its surface gravity. In other words, Eq. (2.18) implies the *non-existence* of **black holes**.

If at first sight it seems paradoxical that a photon cannot be imprisoned gravitationally even though a body of mass $m$ can be, consider that as the photon rises in the gravitational well, it requires less and less energy to continue, both because $r$ is increasing *and* because the effective mass $h\,\nu/c^2$ is decreasing. But for this latter consideration, the photon might not be able to escape. In other words, it continues outward at velocity $c$ by converting "equivalent mass" into potential energy. The loss is always proportional to $m$ and $m$ is therefore never completely consumed.

**(4)  The radar echo delay is the first of the tests to require an actual** measurement of a time interval. Experimentally, this is done by a "molecular" or "atomic" clock. We must therefore determine in advance whether gravitation has any effect upon a molecular clock and if so, what that effect is.

To say that there is a delay is to say that there are more "ticks" on a terrestrial clock than would have been registered by a classical clock. There are two factors which contribute to the delay. First, by Eq. (2.5), there is the factor $e^{2\mu/c^2 r}$. This factor becomes unity in any one of the limits $\mu \to 0$, $c \to \infty$, $r \to \infty$. Thus the classical limit ($c \to \infty$) yields the same result as $r \to \infty$. But the latter describes the behavior of a cosmic molecular or atomic clock keeping time $t_0$ (not to be confused with the present age of the universe in §1). It is therefore independent of the rate of a terrestrial atomic or molecular clock keeping time $t_E$. In the presence of the large mass $M$, the latter is related to the cosmic clock by the equation (2.18), which in this case becomes

$$\nu_E = \nu_0\,e^{\mu/c^2 r}.$$

Therefore, the combination of factors by which the lapse of terrestrial time must be calculated is the cosmic time $r^2 d\theta\,/\,h$ times the photon's comparative rate $e^{2\mu/c^2 r}$ times the relative terrestrial clock rate $e^{\mu/c^2 r_E}$. In sum, therefore, the delay from the earth $(E)$ to sun, or more generally, from any planet $(P)$ to sun will be

$$\Delta t = t_P - t_c = \frac{1}{h} \int \left[ e^{\mu/c^2 (2/r + 1/r_P)} - 1 \right] \frac{d\theta}{u^2}.$$

Let us now evaluate the areal velocity $h$, which, since it is a constant, may be evaluated at any convenient point. Let us choose the point of tangency to the sun, for there $r = R$ and $r\dot{\theta} = c$, giving $h = r \times (r\dot{\theta}) = Rc$. Making this substitution for $h$, and expanding the quantity in brackets into a series, we get

$$\Delta t = \frac{1}{Rc} \int \frac{\mu}{c^2} \left[ 2u + \frac{1}{r_P} \right] \frac{d\theta}{u^2} = \frac{\mu}{Rc^3} \int \left[ \frac{2}{u} + \frac{1}{r_P} \frac{1}{u^2} \right] d\theta.$$

But $\frac{1}{u} \approx R \sec \theta$, so that

$$\Delta t = \frac{\mu}{Rc^3} \int \left[ 2R \sec \theta + \frac{R^2}{r_P} \sec^2 \theta \right] d\theta$$

(2.20)

$$= \frac{\mu}{c^3} \left[ 2\ln\left( \sec \theta + \tan \theta \right) + \frac{R}{r_P} \tan \theta \right]_0^{\theta_P}.$$

Since to better than 1 part in 1000, it is true that $\sec \theta = r_P / R$, this is

$$\Delta t = \frac{\mu}{c^3} \left\{ 2\ln\left( \frac{r_P}{R} + \left[ \left( \frac{r_P}{R} \right)^2 - 1 \right]^{1/2} + \frac{R}{r_P} \left[ \left( \frac{r_P}{R} \right)^2 - 1 \right]^{1/2} \right) \right\}$$

(2.21)

$$\approx \frac{\mu}{c^3} \left[ 2\ln\left( \frac{2r_P}{R} \right) + 1 \right] = 4.926 \left[ 2\ln\left( \frac{2r_P}{R} \right) + 1 \right] \text{ microseconds}.$$

Observations made at Venus' superior conjunction (point of nearest apparent approach to the sun) on January 25, 1970, when the photon's perihelion distance would have been 3.9 R rather than 1.0 R, gave a delay of 190 microseconds rather than the predicted 198 microseconds. However, much more accurate observations at Mars' superior conjunction on November 26, 1976, were made by means of radio signals sent to and reflected from the Viking spacecraft and their landers. These results were accurate to 1 part in 1000 and gave agreement with the equation above. This is truly remarkable, for the delay is only 267 microseconds during a travel time of about 45 minutes!

**(5) Gravitation is a force so omnipresent that it is taken for granted, though** ignored at one's peril. Electromagnetic forces are also commonplace and likewise seldom thought of because familiarity has rendered them routine. These two kinds of forces represent two of the family of forces of nature. They contrast in the fact that gravitation — the interaction of mass upon mass — is invariably an attractive force, never a repulsive one, whereas electromagnetic forces are sometimes those of attraction and sometimes those of repulsion. A distinction of perhaps even greater moment is the fact that electromagnetic forces are enormously stronger than gravitational ones: the electrical attraction between proton and electron is some $10^{39}$ times as great as their mutual gravitational attraction. As a result, electrically neutral matter may aggregate into enormous bodies such as supergiant stars or interstellar molecular clouds, but aggregations of like-charged particles are either of much more limited size or are widely dispersed because of the repulsive forces between particles of the same sign.

General Relativity predicts that gravitationally accelerated bodies will emit gravitational radiation. But if gravitomagnetism exists, why has it not been detected? In addition to the comparative weakness of gravitation relative to electromagnetism, the accelerations of familiar gravitating bodies are also quite small. For example, the orbital acceleration of the earth about the sun is only 2/9 inches per sec per sec, and that of the moon about the earth is only about 1/20 inches per sec per sec. The smallness of the accelerations is not compensated by the magnitude of the velocities; though the earth's orbital velocity is 18.5 miles per sec, this is only $10^{-4}$ the velocity of light.

The possibility of detecting such radiation directly therefore originally seemed not to be feasible because the radiation is so extremely weak. Then, with the discovery of the binary pulsar PSR 1913+16, a method of detecting indirectly its gravitational radiation presented itself. The binary is a pair of mutually revolving neutron stars separated by somewhat more than 4 million miles and revolving in a period of about 8 hours. It appears to be radiating gravitational energy whose loss causes the stars to draw closer together and shortens their period of mutual revolution. Since one of the pair is a **pulsar** — an object from which radio pulses are received every 59 milliseconds — the period of revolution can be determined to within about $2 \times 10^{-5}$ seconds. The decrease in its period of revolution, though very slow, is cumulative and over a period of time can be measured. In this way, it has been found that the stars are in fact spiraling slowly toward each other and at a rate predicted almost exactly by general relativity. In addition, the pulsar's spin axis should precess, much as that of a spinning electron. This too has been observed. This was properly regarded as an observational vindication of the general theory.

By contrast, classical Newtonian theory seemed powerless to predict any gravitational radiation, much less account for the observations of PSR 1913+16. The general theory thus achieved unquestioned dominance both because of its own success and because of the perceived failure of classical theory.

Nevertheless, the two kinds of force have one very important similarity: the attraction of oppositely charged particles at rest and the attraction of mass points obey inverse square laws — Coulomb's and Newton's, respectively. Should we not then expect some similarity in the equations of electromagnetism and gravitation?

At first glance this expectation appears to be confounded, for Newton's law of gravitation bears no similarity to the Lorentz force equation or to Maxwell's equations of electromagnetism. In particular, gravitational theory seems not to have an analogue of magnetism. While on the one hand, we readily acknowledge *electro*magnetism, we do not on the other hand refer to *gravito*magnetism. Can a gravitomagnetic theory of gravitation in flat space-time be derived which will parallel the theory of electromagnetism?[*]

---

[*] "Since the Lorentz transformation is not clever enough to know what the symbols stand for, it follows that the forces and fields of two masses must obey the Maxwell equations in the same way as two opposite charges in electromagnetics." (Petr Beckmann, EinsteinPlusTwo(GolemPress,Boulder,Colorado 1987), p. 166)

In seeking to answer this question, let us review a few of the pertinent equations. The Lorentz equation, for example, is

$$(2.22) \qquad F^{\mu} = m\, a^{\mu} = m\, F^{\mu\nu}\, v_{\nu} \,,$$

where $F^{\mu}$ is the 4-dimensional force vector, $F^{\mu\nu}$ is the electromagnetic tensor, and $v_{\nu}$ is the covariant 4-velocity. Let us identify the kinematic equivalents of the six distinct components of the electromagnetic tensor.

For this purpose, recall that if $v^{\mu} = dx^{\mu}/ds$ is the 4-velocity, then

$$(2.23) \qquad \left( v^{\mu}\, v_{\mu} \right) = 1 \,.$$

Therefore

$$(2.24) \qquad \frac{\delta}{\delta s} \left( v^{\mu}\, v_{\mu} \right) = 2\, a^{\mu}\, v_{\mu} = 0 \,, \quad a^{\mu}\, v_{\mu} = 0 \,.$$

From these results we see that

$$(2.25) \qquad \begin{aligned} \left( \delta^{\mu\nu}_{\alpha\beta}\, a^{\alpha}\, v^{\beta} \right) v_{\nu} &= \left( a^{\mu}\, v^{\nu} - a^{\nu}\, v^{\mu} \right) v_{\nu} \\ &= \left( v^{\nu}\, v_{\nu} \right) a^{\mu} - \left( a^{\nu}\, v_{\nu} \right) v^{\mu} = 1 \cdot a^{\mu} - 0 \cdot v^{\mu} = a^{\mu} \,. \end{aligned}$$

The Lorentz force equation may therefore be written as

$$(2.26) \qquad m \left[ \delta^{\mu\nu}_{\alpha\beta}\, a^{\alpha}\, v^{\beta} - F^{\mu\nu} \right] v_{\nu} = 0 \,.$$

We now invoke the Quotient Law of tensor analysis, which shows the kinematic components of the electromagnetic tensor to be

$$(2.27) \qquad F^{\mu\nu} = \delta^{\mu\nu}_{\alpha\beta}\, a^{\alpha}\, v^{\beta} \,.$$

There is nothing in this equation to identify the accelerations and velocities as being those of electrical attractions (or repulsions!) rather than gravitational attractions. There is therefore nothing to prevent our writing

$$(2.28) \qquad F^{\mu} = m\, a^{\mu} = m\, F^{\mu\nu}\, v_{\nu} = m\, \delta^{\mu\nu}_{\alpha\beta}\, a^{\alpha}\, v^{\beta}\, v_{\nu}$$

as the equations for gravitation. The entire further development of electromagnetism may then be duplicated, including the wave equations, subject only to the caution that no negative masses exist. This implies that gravitational effects propagate through empty space with velocity $c$ just as does light.

**In order to predict gravitational radiation from a pair of mutually gravitating** bodies, we substitute their accelerations and velocities into the relevant equations. We start with Kepler's Law of Ellipses, from which we know that the orbit of a secondary about a primary is

$$(2.29) \qquad r = \frac{a\,(\,1 - e^{2}\,)}{1 + e\cos\theta}$$

in plane polar coordinates. In this orbit, the Law of Areas requires that

$$(2.30) \qquad r^2\,\dot{\theta} \;=\; \frac{2\,\pi\,(\,a\,)^2\,\sqrt{1-e^2}}{P} \;=\; h \;=\; \mathrm{constant}\;.$$

Kepler's Harmonic Law then states that

$$(2.31) \qquad (\,m+M\,)(\,P\,)^2 \;=\; (\,a\,)^3\;,$$

where the masses are in solar units, the period in years, the major semi-axis in astronomical units, and the gravitational constant is $G = 4\,\pi^2$ .

By differentiating Equation (2.31), we see that

$$(2.32) \qquad \dot{P} \;=\; \frac{3\,(\,a\,)^2}{2\,(\,m+M\,)\,P}\,\dot{a}\;;$$

thus the period will decrease if the major semi-axis decreases.

Now from celestial mechanics

$$(2.33) \qquad \dot{a} \;=\; \frac{2\,(\,a^2\,)\,v}{G\,(\,m+M\,)}\,T\;,\quad \therefore\quad \dot{P} \;=\; \frac{3\,(\,a\,)^4\,v}{G\,(\,m+M\,)^2\,P}\,T\;,$$

where $T$ is the tangential component of the perturbing acceleration. This is the rate at which radiation carries off momentum, the energy flux divided by $c$. Thus

$$(2.34) \qquad T \;=\; \frac{\dot{E}(\,m\,)+\dot{E}(\,M\,)}{c}\;.$$

Can $T$ be calculated in a manner analogous to the method employed in electromagnetism? Now although equation (2.28) has been transcribed directly from electromagnetism, there is nothing in it to identify the accelerations as having originated from electrical attractions (or repulsions!) rather than gravitational attractions. There is therefore nothing to prevent our writing

$$(2.35) \qquad F^{\,\mu} \;=\; m\,a^{\,\mu} \;=\; m\,F^{\,\mu\nu}\,v_{\nu}$$

as the equations for gravitation, where the **gravitomagnetic tensor** is (by Eq. (2.28))

$$(2.36) \qquad \delta^{\mu\nu}_{\alpha\beta}\,a^{\,\alpha}\,v^{\,\beta} = F^{\,\mu\nu} = \left\|\begin{array}{cccc} 0 & -c\,H^3 & c\,H^2 & E^1 \\ c\,H^3 & 0 & -c\,H^1 & E^2 \\ -c\,H^2 & c\,H^1 & 0 & E^3 \\ -E^1 & -E^2 & -E^3 & 0 \end{array}\right\|,$$

$$(2.37) \quad = \begin{Vmatrix} 0 & a^1 v^2 - a^2 v^1 & a^1 v^3 - a^3 v^1 & a^1 v^4 - a^4 v^1 \\ a^2 v^1 - a^1 v^2 & 0 & a^2 v^3 - a^3 v^2 & a^2 v^4 - a^4 v^2 \\ a^3 v^1 - a^1 v^3 & a^3 v^2 - a^2 v^3 & 0 & a^3 v^4 - a^4 v^3 \\ -a^1 v^4 + a^4 v^1 & -a^2 v^4 + a^4 v^2 & -a^3 v^4 + a^4 v^3 & 0 \end{Vmatrix}$$

its respective components being the gravitational analogues of the corresponding components of the electromagnetic tensor in Eq. (2.36).

To predict the rate of gravitational radiation, let us first express the gravitomagnetic tensor in terms of the acceleration and velocity of the radiating mass $\boldsymbol{m}$. Since the acceleration and velocity of a secondary about a primary are

$$(2.38) \quad a^i = \left( \ddot{r} - r\dot{\theta}^2 = -\frac{\tilde{G}(m+M)}{r^2}, 0, 0 \right) \text{ and } v^i = (\dot{r}, \dot{\theta}, 0),$$

we have that

$$\mathrm{E}^i(\boldsymbol{m}) = (a^1 v^4 - a^4 v^1, 0, 0) \simeq \left( -\frac{\tilde{G}(m+M)}{r^2}, 0, 0 \right)$$

(2.39)

$$\mathrm{H}^i(\boldsymbol{m}) = \left( 0, 0, \frac{1}{c}[a^2 v^1 - a^1 v^2] \right) = \left( 0, 0, \frac{1}{c}\frac{\tilde{G}(m+M)}{r^2}\dot{\theta} \right)$$

to the lowest order of $1/c$.

We must here take note that these expressions are in terms of the *relative* accelerations whereas in an inertial system they should be *absolute* accelerations. We therefore introduce the factor $\dfrac{\mathrm{M} \text{ or } \boldsymbol{m}}{\boldsymbol{m} + \mathrm{M}}$ as required, giving finally

$$\mathrm{E}^i(\boldsymbol{m}) = \left( -\frac{\tilde{G}\mathrm{M}}{r^2}, 0, 0 \right), \ \mathrm{H}^i(\boldsymbol{m}) = \left( 0, 0, \frac{1}{c}\frac{\tilde{G}\mathrm{M}}{r^2}\dot{\theta} \right)$$

with corresponding expressions for the mass $\mathrm{M}$.

From this we may calculate the Poynting vector for a gravitating mass $\boldsymbol{m}$ as

$$s_i(\boldsymbol{m}) = \boldsymbol{m}\,\varepsilon_{ijk}\mathrm{E}^j\mathrm{H}^k = \boldsymbol{m}\sqrt{g}\,e_{ijk}\mathrm{E}^j\mathrm{H}^k$$

$$(2.40) \qquad = \boldsymbol{m}\left( 0, \frac{r^2}{c}\left[ -\frac{\tilde{G}\mathrm{M}}{r^2} \right]\left[ \frac{\tilde{G}\mathrm{M}}{r^2} \right]\dot{\theta}, 0 \right)$$

$$= \boldsymbol{m}\left( 0, -\frac{1}{c}\left[ \frac{\tilde{G}^2\,\boldsymbol{m}\mathrm{M}^2}{r^4} \right]h, 0 \right)$$

where

$$(2.41) \qquad h = \frac{2\pi(a^2)\sqrt{1-e^2}}{P} \quad \text{and} \quad \tilde{G} = \frac{8\pi G}{c^2}$$

is the gravitational constant appropriate to the number of electromagnetic units of charge which would produce the same gravitational force[*]; it is the same as the constant of proportionality found in the field equation of general relativity, the mediating relation between mass and electric charge. Then, with $\lambda^i = \mathbf{v}^i(m)/|\mathbf{v}^i(m)| = \mathbf{v}^i/\mathbf{v}$ a unit vector along the tangent,

$$(2.42) \qquad \mathbf{T}(m) = s_i(m)\lambda^i = s_i(m)\frac{\mathbf{v}^i}{\mathbf{v}}$$

Unfortunately, the two expressions for the radiations from $m$ and $\mathbf{M}$ separately cannot simply be added as two scalar quantities inasmuch as the Poynting vectors from the separate masses must be added vectorially at a common time and place, constructively and/or destructively, before they are projected upon the tangential direction; only then may they be summed over all directions. Such a calculation would be both lengthy and tedious. We therefore propose an *approximation* to bypass this calculation by introducing a factor $k$ which would compensate so as to give the same result as the full calculation for a pair of electric charges mutually revolving in a circular orbit. This factor is $k = 2/5$. Then

(2.43)

$$[s_i(m) + s_i(\mathbf{M})]\lambda^i = \frac{2}{5}\frac{\mathbf{v}}{c}\left( \frac{16\pi^2 G^2}{c^4 r^4 P} \cdot \frac{m\mathbf{M}^2 + \mathbf{M}m^2}{G(m+\mathbf{M})^2} \cdot \frac{2\pi(a)^2\sqrt{1-e^2}}{P} \right)\frac{\dot{\theta}}{\mathbf{v}} .$$

Combining the various equations, we get finally

$$\dot{P} = -\frac{3a^4\mathbf{v}}{G(m+\mathbf{M})^2 P} \times \frac{2}{5c} \times (m\mathbf{M}^2 + m^2\mathbf{M})\left[ -\frac{\tilde{G}}{r^2} \right]^2 (r^2\dot{\theta})\frac{\dot{\theta}}{\mathbf{v}}$$

$$= -\frac{2}{5c}\frac{3m\mathbf{M}\tilde{G}^2}{G(m+\mathbf{M})P}\left[ \frac{(1-e\cos\theta)^4}{(1-e^2)^4} \right]\left( \frac{2\pi a^2\sqrt{1-e^2}}{P} \right)\dot{\theta} .$$

Substituting the values of $G$ and $\tilde{G}$ gives

$$(2.45) \qquad \dot{P} = -\frac{2^{10}\pi^5}{5c^5}\frac{3m\mathbf{M}}{(m+\mathbf{M})}\left( \frac{a^2}{P^2} \right)\frac{(1+e\cos\theta)^4\dot{\theta}}{(1-e^2)^{7/2}} .$$

---

[*] See Peter Beckmann, *Einstein Plus Two* (Golem Press, Boulder, Colorado, 1987), p. 166; E. T. Whittaker, *A History of the Theories of Aether & Electricity*, (Vol. II, Harper Torch Books, New York) 1960, p. 170.

But from the Harmonic Law (Equation (10)),

(2.46)
$$\frac{a^2}{P^2} = \frac{(m+M)^{2/3}}{P^{2/3}}$$

so that

(2.47)
$$\dot{P} = -\frac{2^{10}\,\pi^5}{5\,c^5}\,\frac{3\,m\,M}{(m+M)^{1/3}\,P^{2/3}}\,\frac{(1+e\cos\theta)^4\,\dot{\theta}}{(1-e^2)^{7/2}}.$$

Since this is a very small quantity which cannot be measured directly, we set

$$\dot{P} = \frac{dP}{dt} \approx \frac{\Delta P}{\Delta t}\quad\text{where}\quad \Delta t = \int_{t_0}^{t_0+P} dt = P$$

and

(2.48)
$$\Delta P = -\frac{2^{10}\,\pi^5}{5\,c^5}\cdot\frac{3\,m\,M}{(m+M)^{1/3}\,P^{2/3}}\cdot\frac{\oint(1+e\cos\theta)^4\,d\theta}{(1-e^2)^{7/2}}.$$

The final result is then

(2.49)
$$\dot{P}(e) = -\frac{2^{11}\,\pi^6}{5\,c^5}\cdot\frac{3\,m\,M}{(m+M)^{1/3}\,P^{5/3}}\cdot\frac{1+3\,e^2+\frac{3}{8}\,e^4}{(1-e^2)^{7/2}}.$$

for the decrease in the period of revolution per revolution. For comparison, the expression given by general relativity is

(2.50)
$$\dot{P}(e) = -\frac{2^{11}\,\pi^6}{5\,c^5}\cdot\frac{3\,m\,M}{(m+M)^{1/3}\,P^{5/3}}\cdot\frac{1+\frac{73}{24}\,e^2+\frac{37}{96}\,e^4}{(1-e^2)^{7/2}}.$$

With $c$ = 63,242 astronomical units/year, $P = 27{,}907\text{ seconds} = 0.00088434$ years, $m$ = 1.386 solar masses, $M = 1.442$ solar masses, $e = 0.617$, the rate of decrease in period predicted by Equation (2.49) is $6.6423 \times 10^{-8}$ seconds per period, within $0.86$ per cent of the observed value of $6.700 \times 10^{-8}$ seconds per period. The value predicted by General Relativity is $6.6948 \times 10^{-8}$ seconds per period. The slight difference between them is to be attributed to the approximation made in by-passing the full calculation for the double point source.

**(6)  Objects of very large mass are deduced to be at the center of mass of all** galaxies and many globular clusters. Their large masses are inferred from the otherwise inexplicably large kinetic energies of objects in the immediate neighborhood of the apparent center of mass, even though no corresponding luminous object can be identified as the massive object itself. It is thus declared by orthodox theory to be a black hole. Interestingly, the mass attributed to the black hole is proportional to the mass of the system of which it is the central object. The properties to be accounted for, therefore, are (1) its invariable centrality, (2) its blackness, (3) its proportionality to the mass of the entire system, and (4) the universality of such objects.

Our own Milky Way Galaxy serves as an example. In a triumph of astrometric technology, a team of astrometrists using high-resolution infrared techniques has determined that the object designated as Sagittarius $\mathbf{A}^*$, a radio source identified as at the mass center of the Milky Way, has a mass of roughly 2.6 million solar masses. This is implied from Kepler's Harmonic Law by the motion of a faint star designated IRS2, whose orbit about $\mathbf{Sgr\ A}^*$ has a major semi-axis of 950 astronomical units and a period of 15 years.

The Schwarzschild radius, also known as the radius of the **event horizon** of such a massive object, is approximately $0.05$ astronomical units. If, however, the massive central object were a neutron star (almost certainly not) whose density is about $10^{15}$ times the density of water, its radius would be roughly $0.003$ astronomical units ($250,000$ miles). The massive object could be even smaller, however, if it were composed of quarks, the constituent particles of protons and neutrons.

Let there be two stars whose baryonic masses are $\mu_1$ and $\mu_2$; these are their masses infinitely far from all other stars or in the classical limit $c \to \infty$. Their gravitational masses are then

$$m_1 = \mu_1\, e^{\frac{G}{c^2}\frac{\mu_2}{r_{12}}} \quad\text{and}\quad m_2 = \mu_2\, e^{\frac{G}{c^2}\frac{\mu_1}{r_{21}}},$$

respectively. Now let a third star with baryonic mass $\mu_3$ be brought from infinity. The gravitational masses of the three stars will then be

$$m_1 = \mu_1 \exp\frac{G}{c^2}\left[\frac{\mu_2}{r_{12}} + \frac{\mu_3}{r_{31}}\right],\ m_2 = \mu_2 \exp\frac{G}{c^2}\left[\frac{\mu_3}{r_{23}} + \frac{\mu_1}{r_{12}}\right],$$

$$m_3 = \mu_3 \exp\frac{G}{c^2}\left[\frac{\mu_1}{r_{31}} + \frac{\mu_2}{r_{23}}\right].$$

The sum of these masses is

$$M_g = m_1 + m_2 + m_3 = \sum_{j=1}^{3} \mu_j \exp\frac{G}{c^2}\sum_{i \neq j}^{3}\frac{\mu_i}{r_{ij}}.$$

Add other stars, one at a time to a total of $N$. Each additional star contributes a term to the exponent of every previous star as well as leads to an additional equation for its own gravitational mass. In general, therefore, the sum of the gravitational masses of the $N$ stars will be

$$M_g = \sum_{j=1}^{N} m_j = \sum_{j=1}^{N} \mu_j \exp\frac{G}{c^2}\left[\sum_{i \neq j}\frac{\mu_i}{r_{ij}}\right] = \sum_{j=1}^{N} \mu_j \exp\left[\frac{G}{c^2}(N-1)\times\overline{\left(\frac{\mu_i}{r_{ij}}\right)}\right]$$

and therefore

$$(2.51) \qquad M_g = N \times \overline{\left[\mu_j \exp(N-1)\frac{G}{c^2}\overline{\left(\frac{\mu_i}{r_{ij}}\right)}\right]}$$

where bars indicate arithmetic means. When $N$ is very large, the difference between $N$ and $N-1$ may be ignored so that finally

$$M_g = N \times \overline{\mu_j \left( \exp N \frac{G}{c^2} \overline{\left( \frac{\mu_i}{r_{ij}} \right)} \right)}.$$

It is hoped that by expressing the gravitational masses in terms of the arithmetic means, one can deal with the case of very large values of $N$ more easily as well as see more clearly the dependence of the gravitational masses upon $N$.

Now it is known from observations that

$$(2.52) \qquad M_g = 6 \times M_b = 6 \sum_{j=1}^{N} \mu_j.$$

With $G = 4\pi^2$ and $c = 63{,}242$ astronomical units per year, this implies that

$$N \frac{G}{c^2} \overline{\left( \frac{\mu_i}{r_{ij}} \right)} = \ln 6 = 1.79 \, , \, N \times \overline{\left( \frac{\mu_i}{r_{ij}} \right)} = 1.79 \times \frac{c^2}{G} = 1.813 \times 10^8.$$

Clearly, if the coefficient of $N$ is not to be infinitesimal, the masses must be intensely concentrated toward the center (as in Sgr A*), perhaps a relic of the Big Bang.

Generalizing Eq. (2.51), the gravitational potential energy of any mass $\mu_i$ will be

$$\mu_j U_j = \frac{c^2}{2} \times \left[ \sum_{i \neq j} m_i \right]^2 = \frac{c^2}{2} \left[ \sum_{i \neq j} 6 \times \mu_i \right]^2 = 18 c^2 \times M_b^2 \ .$$

But the total potential energy of all the stars is

$$\sum_{j=1}^{N} \mu_j U_j = \overline{U} \times \Sigma \mu_j = M_b \times \overline{U}$$

where $\overline{U}$ is the weighted mean of the $U_j$. Therefore

$$\overline{U} = 18 c^2 M_b$$

and $M_b : M_g : \overline{U}/c^2 = 4{:}24{:}72$ per cent, the ratio observed. $M_b$ is the total baryonic mass, $M_g$ the total "dark matter", and $\overline{U}$ the "dark energy".

### 3. Cosmological Problems

**We have at this point laid the necessary groundwork for consideration of** possible model universes, having defined coordinates in space-time and developed a theory of gravitation which accounts for the known major effects. What sort of universe(s) can be imagined which is (are) compatible with these results?

Intuitively, the human mind balks at the notion of a universe of finite extent. If there is a boundary, what is beyond? This is not an illegitimate question. Humorist Stephen Leacock posed the dilemma this way: "We cannot imagine that the stars go on forever. It's unthinkable. But we equally cannot imagine that they come to a stop and that beyond them is nothing, and then more nothing. Unending nothing is as incomprehensible as unending something."

Sir Isaac Newton anticipated the question and offered reasons for his answer. In a letter which he wrote in 1692, he proposed a universe whose general structure he summed up thusly:

> "It seems to me that if the matter of our sun and planets and all the matter of the universe were evenly scattered throughout all the heavens, and every particle had an innate gravity towards all the rest, and the whole space throughout which this matter was scattered was but finite, the matter on the outside of the space would, by its own gravity, tend towards all the matter on the inside, and by consequence fall down into the middle of the whole space, and there compose one great spherical mass. But if the matter was evenly disposed throughout an infinite space it would never convene into one mass; but some of it would convene into one mass and some into another, so as to make an infinite number of great masses scattered at great distances from one another throughout all that infinite space."

Newton's description of a spatially infinite universe filled with matter of uniform density anticipates a model described by Einstein, who wrote in 1921[*]:

> "If we ponder over the question as to how the universe, considered as a whole, is to be regarded, the first answer that suggests itself is surely this: As regards space (and time) the universe is surely infinite. There are stars everywhere, so that the density of matter, although very variable in detail, is nevertheless on the average everywhere the same. In other words: However far we might travel through space we should find everywhere an attenuated swarm of fixed stars of approximately the same kind and density."

Though Einstein discarded this model in favor of the Einstein-deSitter model based on general relativity, let us nevertheless carry out a development of the Newtonian universe in absolute space-time. Such a model, like any credible model, must incorporate as a distinctive feature the cosmological red shift, which was of course unknown to Newton. The cosmological red shift originally established by Edwin Hubble in the late 1920s was assumed as a matter of course to be a Doppler shift. There were a few skeptics, but they offered no credible alternative and strove unsuccessfully against an ever more formidable orthodoxy which today reigns virtually unchallenged.

---

[*]"*The Theory of Relativity*", English translation, Methuen (1921), 4th ed., Ch. 30, p. 105.

Red shifts in galaxies' spectra are related to their presumed recessional velocities $V$ by the special relativistic Doppler shift equation

(3.1)     $$s = \frac{\lambda_0}{\bar{\lambda}} = \left[ \frac{1 + V/c}{1 - V/c} \right]^{1/2}, \quad V = \frac{r}{t} = \text{constant}.$$

To derive this equation, suppose that an observer $O$ reckons that a photon of frequency $\bar{\nu}$ is emitted from a galaxy $\bar{O}$ at a time $t$ (by $O$'s own clock). The clock at $\bar{O}$ would then read a time $\bar{t}$, given by the Lorentz transformation between $O$ and $\bar{O}$ (see Ex. (1.1)) as

$$\bar{t} = \left[ 1 - V^2/c^2 \right]^{1/2} t.$$

Let the photon emitted from $\bar{O}$ arrive at $O$ at time $t_{(0)}$. Then

$$t_{(0)} = t + \frac{r}{c} = \left( 1 + \frac{V}{c} \right) t.$$

Therefore

$$t_{(0)} = \left( 1 + \frac{V}{c} \right) \times \frac{\bar{t}}{\left[ 1 - V^2/c^2 \right]^{1/2}} = \left[ \frac{1 + V/c}{1 - V/c} \right]^{1/2} \bar{t},$$

whence

$$\frac{dt_{(0)}}{d\bar{t}} = \frac{\bar{\nu}}{\nu_0} = \frac{\lambda_0}{\bar{\lambda}} = \left[ \frac{1 + V/c}{1 - V/c} \right]^{1/2} = s,$$

as in Equation (3.1). Since in all cases (only positive $V$) $\bar{\nu} > \nu_0$, this is a Doppler red shift.

Whatever the formal structure of the kinematical or dynamical models, both must give the same picture of the universe or **world picture**. The only quantities directly observable are the red shifts $s$ and the numbers of galaxies $N(s)$ with red shifts less than or equal to $s$. Clearly, we must frame the problem in terms of quantities whose values are common to the two models. These are $s$ and $N(s)$.

Let us relate these conclusions to observation. The observed quantity is the cosmological red shift $1 < s = \frac{\bar{\nu}}{\nu} < \infty$. It is related to the recessional velocity $V$ and this, in turn, to the coordinates $r$ and $t$ by Eq. (3.1) and to coordinate $\rho$ by Eqs. (1.6) and (1.7). To summarize and simplify the results thus far: the cosmological red shift $s$, "expansion velocity" $V$, and absolute distance $\rho$ are related by the equations

(3.2)  $$s = \frac{\bar{\nu}}{\nu_0} = \left[ \frac{1 + V/c}{1 - V/c} \right]^{1/2}, \quad V/c = \tanh \frac{\rho}{ct_{(0)}} = \frac{s^2 - 1}{s^2 + 1}, \quad \rho = c\,t_{(0)} \ln s.$$

Consider, then, a remote galaxy in which a photon of frequency $\overline{\nu}$ is emitted in the frame of reference of an observer $\overline{\mathrm{O}}$ at an absolute distance $\rho$. When it arrives at the earth with frequency $\nu_0$ and red shift $s$ at time $t_0$, it will have been in transit for an absolute time interval $\rho / c$. The absolute epoch of its emission was therefore

$$\overline{\mathrm{T}} = t_{(0)} - \frac{\rho}{c} = t_{(0)} \left( 1 - \ln s \right).$$

We now invoke Eq. (1.8) to show that

(3.3) $$\tau = \frac{t_{(0)}}{s}$$

(3.4) $$t = \tau \cosh \frac{\rho}{c\, t_{(0)}} = \frac{t_{(0)}}{2} \left( 1 + \frac{1}{s^2} \right)$$

(3.5) $$\frac{r}{c} = \tau \sinh \frac{\rho}{c\, t_{(0)}} = \frac{t_{(0)}}{2} \left( 1 - \frac{1}{s^2} \right).$$

Note that $t_{(0)} = \lim\limits_{s \to 1} \left[ t\,(\,s\,) = \dfrac{t_{(0)}}{2} \left( 1 + \dfrac{1}{s^2} \right) \right]$; thus $t_{(0)}$ is determined definitively by the nearest galaxies.

There are interesting and perhaps unforseen implications to Eqs. (3.4) and (3.5). Consider, for example, the values to which $t$ and $r/c$ tend as $s \to \infty$. They are in both cases $t_{(0)}/2$, a somewhat surprising result at first glance, for as $s \to \infty$, $\rho \to \infty$ and $\overline{\mathrm{T}} \to -\infty$. A moment's consideration shows why, however. The most distant galaxies (observers), "receding" with nearly the speed of light, will have receded as much farther in the time since emitting the presently arriving photon as they were when the photon was emitted. Hence the time and distance of the galaxy were only half their present time and distance. However, as one might expect, $\tau \to 0$ and $\overline{t} \to 0$ as $s \to \infty$.

**The Newtonian (dynamical) universe is characterized by uniform density** $D_0$. This means that the number of galaxies in any volume element $d\,\upsilon$ is directly proportional to the volume element itself. Using the fundamental tensor of the spatial portion of the world line element in Eq. (1.9), the volume element is

$$d\,\upsilon = \sqrt{|g_{ij}|}\, dx^1\, dx^2\, dx^3 = (\,c\, t_{(0)}\,)^2 \sinh^2 \frac{\rho}{c\, t_{(0)}} \left[ \cos \varphi\, d\varphi\, d\theta \right] d\rho$$

(3.5) $$= \frac{(\,c\, t_{(0)}\,)^2}{4} \times \left( s - \frac{1}{s} \right)^2 \times \frac{c\, t_{(0)}}{s}\, ds\, d\omega = \frac{(\,c\, t_{(0)}\,)^3}{4} \left( s - \frac{2}{s} + \frac{1}{s^3} \right) ds\, d\omega.$$

Therefore the density of galaxies at epoch $t_{(0)}$ with red shifts between $s$ and $s + ds$ is

(3.6)

$$d\mathrm{N}(s) = \mathrm{D}_0 \frac{(c\,t_{(0)})^3}{4} \left( s - \frac{2}{s} + \frac{1}{s^3} \right) ds\,d\omega$$

$$\mathrm{N}(s) = \mathrm{B} \left( s^2 - 4\ln s - \frac{1}{s^2} \right),$$

where $\mathrm{N}(s)$ is the total number of galaxies with red shifts less than or equal to $s$, $\omega$ is the solid angle within which the count is made, and $\mathrm{B} = \mathrm{D}_0 \dfrac{(c\,t_{(0)})^3}{4}\,\omega$ is a constant to be determined from observation. This model of a Newtonian universe will stand or fall on these results. It should be noted, however, that the function $\mathrm{N}(s)$ represents the upper envelope of the observations inasmuch as galaxies have a range of luminosities and those of low luminosity may fail to be counted when the values of $s$ are large.

It is also to be noted that these results would be the same for the Milne (kinematic) model, for they include no parameters peculiar to either the Newtonian or Milne models. In fact, Milne derived the same equation for his model by an entirely independent argument[*] and showed as well that the distribution of galaxies satisfied the equation of continuity. Fig. 95 is a graph of the function $\mathrm{N}(s)$.

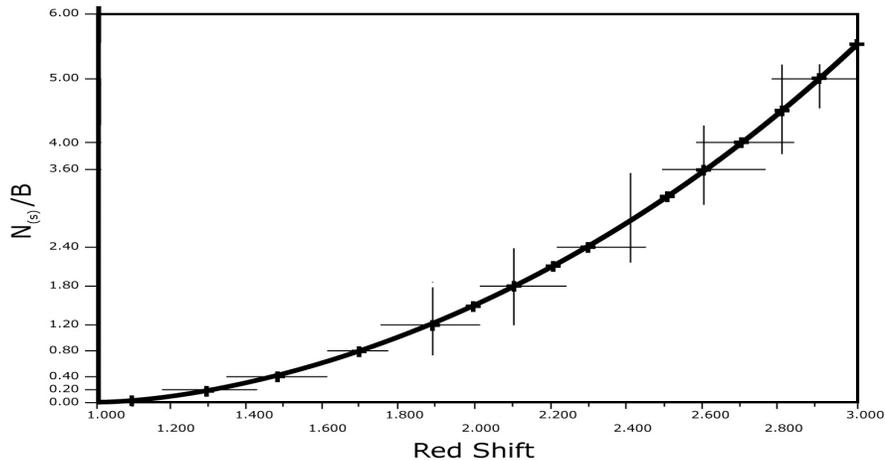

Figure 95

The value of $\mathrm{B}$ may be calibrated by observing the count of the total number of galaxies $\mathrm{N}(s_0)$ with red shifts less than $s_0$ for some conveniently chosen $s_0$. The equation will then predict the number $\mathrm{N}(s)$ for all other red shifts according to this model. Though the ultimate observational test is pending, Milne's models of the universe may even now be compared with current models of general relativity. Let us summarize the results as presented by Milne[**]:

Though there are literally infinitely many general relativistic models defined by a continuum of values of the mean density of matter, they may be grouped for simplicity into three categories – elliptical, parabolic and hyperbolic. These designations correspond to the fact that the trajectories of the galaxies are (1) degenerate ellipses, (2) degenerate parabolae, or (3) degenerate hyperbolae.

The first is also known as the "oscillating universe" inasmuch as its history is an endless series of expansions and subsequent contractions; we are currently in an expanding phase, the Big Bang, but must look forward to an eventual Big Crunch, followed by the next Big Bang. The second is the Einstein-deSitter universe characterized by an eternal expansion to a limiting state in which all galaxies tend to a common velocity zero. The third is a universe in which the galaxies mutually recede, decelerating forever to some positive limiting velocity. These respective behaviors all imply a universe of finite mass, and therefore are inconsistent with the Copernican Principle.

Such finite universes should also in principle be transparent; it should be possible to see the last galaxy (or particle) in any direction. That is, their boundaries should be visible. Here one should be able to witness a most unusual phenomenon: the creation of matter during the stage of expansion and annihilation during the contraction of an oscillating universe. The Newtonian universe was created once, in the infinite past. The Milne universe was created once at time $t_0 = 0$, hidden now behind the curtain of the cosmological background radiation and therefore also invisible. Creation no longer occurs and is therefore not available to inspection even in principle.

All three types of general relativistic universes allow for the existence of a "cosmological constant", a coefficient in the field equations of general relativity whose value must be deduced from observation and may in principle be positive, negative or zero. At present, a positive value is more in vogue than a negative one, though a zero value has not been definitively ruled out. The physical interpretation of a positive cosmological constant, numerically very small, is that masses exert upon each other a repulsive force which increases directly with distance. While the familiar gravitational force is an attraction, the cosmological force is a repulsion. Because of the extreme smallness of the postulated cosmological constant, the attraction overwhelms the repulsion except when the separation of the masses reduces the inverse square attraction almost to zero. Thereafter, the repulsion dominates and for bodies sufficiently far apart, the net "force of gravity" is a repulsion increasing with distance; the farther masses are from each other, the greater their effect upon each other. Clearly, this is a counterintuitive feature of the theory.

There is an interesting related matter involving the red shift. This is the celebrated **Olbers' paradox**. It is a seeming contradiction pointed out in 1826 by Dr. Friedrich Wilhelm Olbers, a Hamburg physician and distinguished amateur astronomer (there were very few professional astronomers in that day); he was the discoverer of several comets which bear his name. Dr. Olbers drew attention to the commonplace fact that the night sky is dark. He argued that if the universe were infinite in extent, homogeneous in composition, static, and populated with stars which have always shone as they presently do, then one's line of sight in any direction whatever would necessarily be intercepted by some star and therefore our entire sky should have the same surface brightness as the average surface brightness of the stars. Since this is decidedly not the case, Dr. Olbers concluded that at least one of his premises was false. Which one or ones?

The resolution of the paradox is somewhat subtle. In short, more distant parts of the universe contribute ever-diminishing increments of visible starlight because of the increase of the red shift with distance, and since the stars and galaxies have not always existed and shone with their present luminosities, we "see through" them to the cosmic background, a region of space whose radiations are so red-shifted that it provides virtually no visible light. Significantly, this implies, that the organized matter of the universe, regardless of model, is of limited age even though in the dynamical model the universe itself is infinitely old.

| $s$ | $V/c$ | A |
|-----|-------|-------|
| 1.0 | 0.000 | 1.000 |
| 1.1 | 0.095 | 0.995 |
| 1.2 | 0.180 | 0.984 |
| 1.3 | 0.257 | 0.967 |
| 1.4 | 0.324 | 0.946 |
| 1.5 | 0.385 | 0.923 |
| 1.6 | 0.438 | 0.899 |
| 1.7 | 0.486 | 0.874 |
| 1.8 | 0.528 | 0.849 |
| 1.9 | 0.566 | 0.824 |
| 2.0 | 0.600 | 0.800 |
| 2.2 | 0.658 | 0.753 |
| 2.4 | 0.704 | 0.710 |
| 2.6 | 0.742 | 0.670 |
| 2.8 | 0.774 | 0.633 |
| 3.0 | 0.800 | 0.600 |
| 3.2 | 0.822 | 0.569 |
| 3.4 | 0.841 | 0.541 |
| 3.6 | 0.857 | 0.516 |
| 3.8 | 0.870 | 0.492 |
| 4.0 | 0.882 | 0.471 |

The major cause of the darkness of the night sky, however, is an effect called **attenuation**. As we have seen, the relation between the time $t$ of a clock at an observer $O$ and the time $\bar{t}$ read from a receding clock at $\bar{O}$ is

$$\bar{t} = \sqrt{1 - V^2/c^2}\, t = \frac{2s}{s^2 + 1}\, t \,.$$

From this we see that

$$\Delta \bar{t} = \frac{2s}{s^2 + 1}\, \Delta t \,.$$

Because $\bar{O}$ is receding from $O$ at a velocity $V$, the radiations emitted at the end of the time interval $\Delta \bar{t}$ will have farther to travel than those emitted at the beginning, and therefore will not have arrived by the expiration of the time interval $\Delta t$. Hence the brightness of the source at $\bar{O}$ will be reduced at $O$ by the factor

$$A = \frac{2s}{s^2 + 1} \,,$$

the attenuation. As the table shows, the attenuation A decreases significantly even for the nearer parts of the visible universe. At the distance of the cosmic background radiation, for which $s \approx 2000$, its radiation is reduced by a factor of $0.001$ or $7.5$ magnitudes! More distant sources, if any, would suffer even greater dimming by attenuation.

As a sidelight, consider the **Hubble constant**, whose value has been the subject of much effort over the years. The Hubble constant is the rate at which the velocity of expansion in the Milne universe increases with distance, namely

$$H = \frac{d\,V}{d\,r} = \frac{d\,V/d\,s}{d\,r/d\,s} = \frac{4}{t_{(0)}}\,\frac{1}{1 + \dfrac{2}{s^2} + \dfrac{1}{s^4}},$$

which ranges from $1/t_{(0)}$ to $4/t_{(0)}$. Clearly, the Hubble constant $H$ (in km/sec/mpsc) is not constant as $s$ goes from $1$ to $\infty$.

| $s$ | $d$ | $H$ | $D$ |
|-----|------|---------|-------|
| 1.0 | 0 | 54.000 | 0.000 |
| 1.1 | 1.189 | 64.750 | 1.192 |
| 1.2 | 2.093 | 75.231 | 2.110 |
| 1.3 | 2.797 | 85.256 | 2.845 |
| 1.4 | 3.355 | 94.707 | 3.450 |
| 1.5 | 3.806 | 103.527 | 3.961 |
| 1.6 | 4.174 | 111.695 | 4.403 |
| 1.7 | 4.480 | 119.220 | 4.792 |
| 1.8 | 4.736 | 126.128 | 5.140 |
| 1.9 | 4.952 | 132.454 | 5.455 |
| 2.0 | 5.138 | 138.240 | 5.744 |
| 2.2 | 5.435 | 148.361 | 6.261 |
| 2.4 | 5.661 | 156.821 | 6.718 |
| 2.6 | 5.837 | 163.917 | 7.130 |
| 2.8 | 5.976 | 169.895 | 7.509 |
| 3.0 | 6.089 | 174.960 | 7.861 |
| 3.2 | 6.181 | 179.276 | 8.191 |
| 3.4 | 6.257 | 182.974 | 8.504 |
| 3.6 | 6.321 | 186.163 | 8.802 |
| 3.8 | 6.376 | 188.927 | 9.087 |
| 4.0 | 6.422 | 191.336 | 9.361 |

The observational determination of the value of the Hubble "constant" requires that the distances of remote galaxies be determined accurately. These distances are generally deduced from objects termed "standard candles", sources whose luminosities are known reliably by independent means. Their observed brightnesses in distant galaxies then imply their distances. However, these apparent brightnesses are less for sources receding rapidly than they would be for stationary sources at the same distance. This is because the radiations from the receding sources must travel ever farther in succeeding seconds and are therefore "strung out" or attenuated. The attenuation is given by the relation

$$d\overline{t} = \sqrt{1 - \frac{V^2}{c^2}}\,dt = \frac{2s}{s^2+1}\,dt,$$

where $dt$ is a small time interval at the earth during which is received radiation emitted by the distant rapidly receding source in time $d\overline{t}$.

Therefore the attenuation must be allowed for else the calculated distance will be greater than the true distance by a factor of

$$\sqrt{\frac{s^2 + 1}{2s}}$$

Column $D$ of the Table shows how much greater the uncorrected distances $D$ would be than the true distances $d$.

This same effect is diminishing the apparent brightnesses of the type Ia supernovae, the most distant standard candles presently observed. Uncorrected distances are then interpreted to imply accelerating expansion of the universe. Presumably the corrected distances would not show this artificial acceleration.

## 4. Model Universes

**M**odel universes are maps in space-time, incorporating features of the observed universe and subject to falsifiability by observation. As with maps of the earth, there may be more than one representation of the universe though the universe itself is unique. What general requirements should a model universe satisfy? First of all, we may ask: What kind(s) of space(s) is (are) suitable for mapping the universe onto? Just as one would not be well-advised to attempt to map the surface of the earth without first having some notion of what geometry is most compatible with the known rules of terrestrial trigonometry, so also one would be well-advised to select space(s) compatible with known characteristics of the observed universe. Are there any specific guidelines for the entire universe?

As previously noted, there are two theorems from differential geometry which recommend themselves. (1) Helmholtz' Theorem declares that **the free mobility of rigid bodies of non-infinitesimal extent requires that the geometry of space be Riemannian and of constant curvature.** Perhaps the bodies which we regard as rigid are not really so, but this is a matter which we cannot assert with assurance. In any case, the presumption in favor of Riemannian space of constant curvature is reinforced by Schur's Theorem which states that (2) **a space can be isotropic only if its curvature invariant is constant.** Since the universe on the largest scale, that of the primeval fireball, appears to be highly (if not precisely) isotropic, it would seem to be reasonable to make it an initial feature of the large-scale universe that it be Riemannian of constant curvature. The most refined observations of the cosmic background radiation to date are interpreted to imply that space-time is indeed flat.

A consideration of a quite different kind is what has been called the Copernican Principle. This principle states that **It is not possible to locate a unique "center" of the large-scale universe.** This notion commends itself on the historical grounds that all attempts to identify the unique center of the universe – the Temple at Jerusalem, the center of the earth, the center of the sun, the center of the Milky Way galaxy – have thus far failed.

If one accepts the Copernican Principle, then the universe must have infinite mass if not necessarily infinite extent, for a finite number of galaxies (or particles) not infinitely dispersed and mutually isolated must have a boundary and a unique center of mass. This would mean that calculation of the mass of any model universe must not yield a finite quantity.

The Newtonian and Milne universes are both mapped onto a 4-space of zero space-time curvature and therefore satisfy the Helmholtz Theorem and Schur's lemma. They likewise satisfy the Copernican Principle in having infinitely many galaxies. From any location in either of these universes, the universe appears to have infinitely many galaxies distributed istropically about the observer. Either may be mapped into the other. The respective mappings are given by two common parameters, the red shift $s$ and the "age of the universe" $t_{(0)}$. The properties of these models are summarized in the Table below.

These two models are of the same universe, but constitute different maps or descriptions of it, just as a terrestrial globe and a Mercator map constitute different descriptions of the same earth's surface. They have one important feature in common, however: their space-time curvature is zero. Now "space-time curvature" is identical with "acceleration"; wherever the phrase "space-time curvature" occurs in general relativistic descriptions of phenomena, one may substitute the single term "acceleration". This conveniently simplifies the description and allows a more ready grasp of the phenomena. Zero space-time curvature therefore is merely an alternate way of describing a universe in which no part is accelerated with respect to any other part.

The two universes differ, however, in the structure they appear to have. Just as a Mercator projection of the earth's surface extends to infinity in the poleward direction, the Newtonian universe extends to infinity in the outward direction. On the globe of the earth, however, the earth's surface extends only to the pole; in the Milne model, the universe extends only to the boundary at $R = c\,t_{(0)}$. In both universes, the density of galaxies (particles) is given by Eq. (3.6). The Milne universe has a creation event at $t = 0$, $t_{(0)}$ years ago. The Newtonian universe has existed forever on the absolute time scale. Which is correct? Both; the choice is arbitrary and depends upon which features of each are deemed the more essential.

| Newtonian Universe | Milne Universe |
|---|---|
| $T(s) = t_{(0)}(1 - \ln s)$ | $t(s) = \dfrac{t_{(0)}}{2}\left(1 + \dfrac{1}{s^2}\right)$ |
| $\rho(s) = c\,t_{(0)} \ln s$ | $r(s) = \dfrac{1}{2}c\,t_{(0)}\left(1 - \dfrac{1}{s^2}\right)$ |
| Spatial curvature: $K = -1/(c\,t_{(0)})^2$ | Spatial curvature: $K = 0$ |
| Age: Infinite | Age: $t_{(0)} \approx 13 \times 10^9$ years |

Note that both the Newtonian and the Milne models expand linearly with time since both $r$ and $\rho$ are proportional to $t_{(0)}$, the "age" of the universe. It is somewhat of an oddity that the infinite Newtonian universe is expanding as well as the finite Milne universe. Clearly, both universes satisfy the Copernican Principle.

Another feature of both the Milne and Newtonian universes comes to light from the common expression for the equation for the expansion velocity in terms of the red shift (Eq. 3.2). By differentiating,

$$\frac{d\mathrm{V}}{ds} = \frac{4\,s\,c}{(\,s^2 + 1\,)^2}.$$

If one seeks to give meaning to this result, one might say that since

$$\left(\frac{d\mathrm{V}}{ds}\right)_{s=\infty} = 0 \ \text{ and } \ \left(\frac{d\mathrm{V}}{ds}\right)_{s=1} = c,$$

the velocity of a young, remote galaxy (large $s$) increased more slowly in the past than in later times (small $s$) and that therefore the expanding universe is accelerating in its expansion. This would, however, be a false conclusion, since

$$\frac{d\mathrm{V}}{dt} = \frac{d\mathrm{V}}{ds} \times \frac{ds}{dt} \ \text{ and } \ \frac{ds}{dt} = 0.$$

The seeming acceleration is an **artifact of the model**, not a property of the expanding universe. Galaxies do not assume the velocities of currently more distant ones as they later arrive at those greater distances.

## 5. The Cosmological Red Shift and the Pure Numbers of Nature

**T**he cosmological red shift has an evident explanation in the kinematic model; it is the Doppler shift produced by the expansion of the universe following the "Big Bang". In the dynamical model, the expansion velocity becomes merely a parameter related to distance. Since every event, history and phenomenon in the expanding universe should have a counterpart in the dynamical universe, what "explanation" accounts for the cosmological red shift in the Newtonian model?

To answer this question requires that we look searchingly at the two methods of time-keeping — the kinematic and the dynamical clocks — in the expanding and the static universes. It is the distinction between the two time scales which lies at the root of the two descriptions of the universe.

Consider an observer $\mathrm{O}$ at the origin. He observes a galaxy $\overline{\mathrm{O}}$ at distance $\overline{r}$ or $\overline{\rho}$ where the time is $\overline{t}$ or $\overline{\mathrm{T}}$. The relation between the coordinates of $\mathrm{O}$ and $\overline{\mathrm{O}}$ is given by the Lorentz transformation. Hence, whereas $\mathrm{O}$ ascribes to an event at $\overline{\mathrm{O}}$ the kinematic time $t$, $\overline{\mathrm{O}}$ himself will read from his own kinematic clock a time

(5.1)
$$\bar{t} = \sqrt{1 - V^2/c^2}\, t .$$

Two events at $\bar{O}$ which succeed each other by a short time $dt$ according to $O$'s clock, will be separated by an interval

$$d\bar{t} = \sqrt{1 - V^2/c^2}\, dt$$

according to $\bar{O}$'s clock. If $\bar{O}$'s red shift is $s$, then

$$\frac{V}{c} = \frac{s^2 - 1}{s^2 + 1}, \quad \sqrt{1 - V^2/c^2} = \frac{2s}{s^2 + 1},$$

whence

(5.2)
$$d\bar{t} = \frac{2s}{s^2 + 1}\, dt .$$

At the same time, since $\bar{O}$'s absolute distance $\rho$ is fixed, the relation between $O$'s determination of $dt$ at $\bar{O}$ and his determination of the corresponding absolute time interval $d\mathrm{T}$ will be

$$dt = \frac{\partial t}{\partial \mathrm{T}}\, d\mathrm{T} = \frac{t}{t_{(0)}}\, d\mathrm{T} = \frac{s^2 + 1}{2s^2}\, d\mathrm{T} .$$

Hence $\bar{O}$'s determination of the interval $d\bar{t}$ (by his own kinematic clock) is related to $O$'s determination of the corresponding interval $d\mathrm{T}$ (by his own dynamical clock) by combining the two previous equations to give

$$d\bar{t} = \frac{2s}{s^2 + 1} \times \frac{s^2 + 1}{2s^2}\, d\mathrm{T} = \frac{1}{s}\, d\mathrm{T} .$$

Evidently, the kinematic clock at $\bar{O}$, keeping kinematic time $\bar{t}$, is running $1/s$ as fast as the dynamical clock at $O$, keeping dynamical time $\mathrm{T}$.

Now since $\mathrm{T}$ and $\bar{\mathrm{T}}$ determined by $O$ and $\bar{O}$ are both absolute times, it must be true that $d\mathrm{T} = d\bar{\mathrm{T}}$. Hence

$$\frac{d\bar{r}}{c} = dt = \frac{1}{s}\, d\mathrm{T} = \frac{1}{s}\, d\bar{\mathrm{T}} .$$

But $d\bar{\mathrm{T}} = \frac{d\bar{\rho}}{c}$ so that $\frac{d\bar{r}}{c} = \frac{1}{s} \frac{d\bar{\rho}}{c}$ since the absolute distance increments $d\rho$ and $d\bar{\rho}$ must also be equal. As a result, it follows that

(5.3)
$$d\bar{r} = \frac{1}{s}\, d\bar{\rho} .$$

In other words, $\bar{O}$'s determination of the kinematic distance increment $d\bar{r}$ is that it is $1/s$ as great as his determination of the corresponding absolute distance increment $d\bar{\rho}$.

Clearly, the kinematic time and distance measurements by $O$ differ from the absolute (dynamical) time and distance measurements of the same quantities by $\bar{O}$ by a common factor $1/s$. Quite generally, $O$'s measurements of lengths and time increments are less than $\bar{O}$'s ascriptions of the values of the same lengths and time intervals. However, a kinematic determination of their *ratio* (a velocity) $d\bar{r}/d\bar{t}$ by $\bar{O}$ gives the same result as a dynamical determination of the same ratio (velocity) $d\bar{\rho}/d\mathrm{T}$ by $\bar{O}$ using absolute time and space coordinates. By extension, the squares of velocities must also be equal and therefore kinetic energies. Because kinetic energies are transformable into energies of other kinds, it follows that measurements of energies by $\bar{O}$ in both systems of time-keeping will in all cases give identical results.

It is a corollary of this argument that quantities which are functions of length and time intervals *not* homogeneous and of order zero will have different values in kinematic and dynamical systems at $\bar{O}$; only when $s = 1$ ($V = 0$, $\rho = 0$, the observer $O$ at the origin) will such quantities have the same value in both kinematic and dynamical systems. This is because we have calibrated the transformation between kinematic and dynamical systems so that $t = \mathrm{T} = t_{(0)}$ and $dt = d\mathrm{T}$ at the present time at $O$, where $s = 1$. At $\bar{O}$, a length $\lambda_0$ on the kinematic scale is a length $\lambda = s\,\lambda_0$ on the dynamical scale and a duration $\theta_0$ on the kinematic scale is a duration $\theta = s\,\theta_0$ on the dynamical scale.

What, therefore, of quantities which depend in dissimilar ways upon the powers of $dr$ and $dt$ (time interval)? If the dimensions of a quantity are such that the factors $s$ do not cancel, then, like $dr$ alone or $dt$ alone, they would be found to have values which diverge progressively as one considers times and distances ever farther from the here and now. This expectation has far-reaching implications.

Consider, for example, Planck's constant $h$, the quantum of action. Since action has the dimension $\mathrm{M}\,\mathrm{L}^2\,\mathrm{T}^{-1}$, the time and distance factors do *not* offset at different epochs, so that

$$\lambda^2\,\theta^{-1}\,\bar{h} = (s^2\,\lambda_0)(s\,\theta_0)^{-1}\,\bar{h}_0 = (s\,\bar{h}_0)\,\lambda_0^2\,\theta_0^{-1}$$

whence

(5.4) $$\bar{h} = s\,\bar{h}_0.$$

That is to say that $O$ will determine that $\bar{h}$ measured on $\bar{O}$'s dynamical scale is $s$ times the value $\bar{h}_0$ measured on $\bar{O}$'s kinematic scale. At the same time, he will find that an atomic frequency is $\bar{\nu} = \dfrac{1}{s}\,\bar{\nu}_0$ since frequency has the dimensions $\mathrm{T}^{-1}$.

The bearing which this has upon the cosmological red shift becomes evident in the relation $\mathrm{E} = h\,\nu$ for the energy of a photon. Since energy is of the dimensions of $\mathrm{M}\,\mathrm{L}^2\,\mathrm{T}^{-2}$, its value remains the same for all times and places. Therefore

$$\bar{h}_0\,\bar{\nu}_0 = \bar{h}\,\bar{\nu} = (s\,\bar{h}_0)\left(\frac{1}{s}\,\bar{\nu}_0\right).$$

This is to say that in a distant galaxy, the frequency $\bar{\nu}$ of the emitted photon is less on the dynamical time scale than the frequency on the kinematic time scale by the factor $s$. This comes about because we are counting atomic vibrations during unequal time increments $d\bar{t}$ and $d\bar{T}$. We therefore could say that the cosmological red shift in the dynamical (Newtonian) model universe is caused either by the greater value of Planck's constant $\bar{h}$ or by saying that a kinematical second was much shorter than a dynamical second in remote past times. The number of atomic vibrations in a "short" second will obviously be fewer than the number of atomic vibrations in a "full" second.

This line of reasoning has further implications of interest when one considers what are called **the pure numbers of nature**. One of these, for example, is the **fine structure constant**

$$(5.5) \qquad\qquad \alpha = \frac{2\pi\kappa e^2}{hc} \approx \frac{1}{137},$$

which is a dimensionless quantity formed of several physical constants whose values can be currently measured by direct means in a terrestrial laboratory. Since it is a dimensionless quantity, its value (like the epoch independence of energy $h\nu$ and the velocity of light $c$) is independent of the system of units. Whether the units are *cgs*, *mks*, British or anything else, the value of $\alpha$ will be the same, independent of time or place. Therefore, since $c$ is epoch-independent in both Newtonian and Milne models of the universe but Planck's constant $h$ is not, we must conclude that $\kappa e^2$ varies in the same way as $h$, i. e.

$$(5.6) \qquad\qquad (\bar{\kappa}\,\bar{e}^2) = s(\kappa e^2).$$

This leads to yet another relation, one based upon the pure number which is the ratio of the electrostatic force between electron and proton and the gravitational force between them. Thus

$$\frac{F_e}{F_g} = \frac{\kappa e^2}{G\,m_p\,m_e}, \quad \text{whence} \quad \bar{G}\bar{m}_p\,\bar{m}_e = s\left(G\,m_p\,m_e\right).$$

Since masses are epoch-independent, being determined from ratios of velocities, we conclude that

$$(5.7) \qquad\qquad \bar{G} = s\,G.$$

Do these differences in $h$, $\kappa$ and $G$ provide means for discriminating between the two models? Can observation indicate which is the "true" model of the universe? The answer must be "No", for no transformation of coordinates can alter the phenomena, only the description of those phenomena. The observations can discriminate only which time scales have been used, explicitly or implicitly. The kinematic and dynamical time scales are rooted in observational procedures, are locally indistinguishable, and provide alternative descriptions of phenomena which, though fundamentally different, are not outrageously inconvenient or implausible. Aspects of each are intuitively appealing.

**Two important points remain. The first is consideration of the primeval** fireball radiation. In 1965, the presence of this blackbody radiation of effective temperature 2.726°K was first detected. If the earth's peculiar velocity (i.e., velocity characteristic only of the earth) is allowed for, the primeval fireball radiation is isotropic to less than the present limits of detection, i.e., to within presumptive temperature variations of a very small fraction of a degree Kelvin. In the kinematic model, the fireball radiation's temperature is generally taken to be a mere $2.726°\text{K}$ because the radiations have been velocity red-shifted by a factor $s \approx 2000$. In the dynamical model, however, one would say simply that $2.726°\text{K}$ is the temperature of unorganized matter at the great distance and much earlier epoch of the fireball. These are merely alternative ways of explaining the same observations.

One body of opinion holds that the apparently rigorous isotropy of the fireball radiation constitutes an observable property of the universe which requires accounting for. This presupposes that in the kinematic model the universe had a least age $t = \Theta$, a very small value such as $10^{-38}$ seconds. The particular value of $\Theta$ is of less moment than the fact that $\Theta > 0$. Were this the case, diametrically opposite parts of the universe would have been separated by a distance $\Delta = 2c\Theta$ at the event of "creation" and would have forever remained mutually inaccessible since not even light could overtake an opposite point receding from it at the speed of light. This impossibility of any interaction between them would render highly unlikely their having identical states long afterward, and would call for some deeper analysis of the isotropy of the primeval fireball radiation.

The result has been a recent model known as the **inflationary universe**. The theory of the inflationary universe is extremely abstruse. It is grounded in the multi-dimensional physics of high-energy particles. Moreover, the fireball itself affords an impenetrable opaque screen through which such early states of the universe cannot be seen. All matter at its inner boundary is and always has been in interaction with all other such matter.

A final word needs to be offered concerning the "constant" $t_{(0)}$. It has been interpreted as "the present age of the universe". Since that present age is advancing, a cosmologist at a much earlier epoch would have used a much smaller value of $t_{(0)}$. He would still have found $T$ to range from $-\infty$ to $+\infty$ and $\tau$ from $0$ to $+\infty$, and the lower observable limit of $t$ would in any case have been $t_{(0)}/2$, then a correspondingly smaller lower limit. We might say, therefore, that $t_{(0)}$ is a "time-dependent constant". It has a fixed value for all present calculations, but for convenience' sake is adjusted as the universe ages. The epoch of observation is therefore a parameter of the universe.

## Appendix

The major semi-axis of a binary orbit is related to the position and velocity of the secondary by the equation

$$a = \frac{G(m+M)}{\dot{r}^2 + r^2\dot{\theta}^2 - \dfrac{2G(m+M)}{r}}.$$

Therefore $\quad \dfrac{da}{dt} = \dot{a} = 2\,\dfrac{a^2}{G(m+M)}\left[\dot{r}\ddot{r} + r\dot{r}\dot{\theta}^2 + r^2\dot{\theta}\ddot{\theta} + \dfrac{G(m+M)\dot{r}}{r^2}\right].$

Now let $p^i$ be a perturbing acceleration. Then, eliminating $\ddot{r}$ and $\ddot{\theta}$,

$$\frac{G(m+M)\dot{a}}{2a^2} = \dot{r}\left[r\dot{\theta}^2 - \frac{G(m+M)}{r^2} + p^1\right] + r\dot{r}\dot{\theta}^2$$

$$+ r^2\dot{\theta}\left[-\frac{\dot{r}\dot{\theta}}{r} + p^2\right] + \frac{G(m+M)\dot{r}}{r^2}$$

$$= \dot{r}p^1 + \left[r^2\dot{\theta}\right]p^2 = v_1 p^1 + v_2 p^2 = v_i p^i = T.$$

But $v_i p^i$ is the projection of the perturbing acceleration upon the tangent, hence equal to the magnitude of the velocity times the tangential perturbing acceleration $T$. In other words,

$$\frac{da}{dt} = \frac{2a^2 v}{G(m+M)}\, T.$$

## Notes

**§1.** The Lorentz transformation has been given an extremely elegant derivation in Whittaker (23). He first derives the velocity addition formula (§21, pp. 49—51), later using this to obtain the Lorentz transformation (§26, pp. 60—64). A thoroughgoing general analysis of time-keeping and its relation to equivalences and congruences is to be found in Milne (9), Part I, Chs. I-IV. There is included also a unique derivation of the Lorentz transformation.

**§2.** The Lagrangian function (Eq. 2.2) appears to have been first given by A. G. Walker in *Nature*, v. 168 (1951), pp. 961-962. It is there stated but not derived. Walker claims for it only the advance of Mercury's perihelion and the bending of light at the limb of the sun. This is a seminal Note.

Will (24) has given an engaging popular account of the theory of general relativity and the various tests to which it has been put. He also discusses (Ch. 9) the "constancy" of the gravitational constant (see §4, this chapter). The bibliography (pp. 259—261) is especially helpful for the general reader. Beckmann (2) provides many fascinating historical asides to the development of relativity theory as well as unorthodox derivations of many of its results.

**§5.** The pure numbers of nature have an intrinsic interest quite apart from their property of being dimensionless. The most significant of them, combining physical constants from microphysics to cosmology, cluster about a few numbers of the order of the zeroth, 39th and 78th powers of 10. A chance distribution of this particular sort is almost inconceivably improbable. Whittaker's book (23) is an absorbing discussion of Eddington's attempt to derive these pure numbers exactly from fundamental theory, including the theory of relativity, of which he was one of the earliest authorities and most masterful expositors. Bondi (3) also discusses the pure numbers of nature in Chapter VII.

# Bibliography

Many fine books are available on vector and tensor analysis, differential geometry, classical and relativistic physics, etc. The following works have been selected as most likely to be helpful in supplementing the text on closely related fields or in giving alternate treatments of common topics.